\pdfoutput=1
\documentclass[fleqn,usenatbib,usedcolumn]{mnras}

\usepackage{newtxtext}
\usepackage{newtxmath}

\usepackage[T1]{fontenc}

\let\la=\lesssim

\usepackage{graphicx}	
\usepackage{amsmath}	
\usepackage{bm}		
\usepackage{mathrsfs}
\usepackage{ctable}

\newcommand{\angstrom}{\mbox{\normalfont\AA}}

\title[Quasar Lifetimes at $z \simeq 4$]{Evidence for Short $\sim 1\,{\rm Myr}$ Lifetimes from the \ion{He}{ii} Proximity Zones of $z \sim 4$ Quasars}

\author[I.S. Khrykin et al.]{
Ilya S. Khrykin,$^{1}$\thanks{E-mail: i.khrykin@gmail.com}
Joseph F. Hennawi,$^{2,3}$
G\'abor Worseck$^{4}$
\\
$^{1}$Southern Federal University, Stachki Avenue 194, 344090, Rostov-on-Don, Russia\\
$^{2}$Department of Physics, University of California, Santa Barbara, CA 93106, USA\\
$^{3}$Max-Planck-Institut f\"ur Astronomie, K\"onigstuhl 17, D-69117 Heidelberg, Germany\\
$^{4}$Institut f\"ur Physik und Astronomie, Universit\"at Potsdam, Karl-Liebknecht-Str.\ 24/25, D-14476 Potsdam, Germany
}

\date{Accepted XXX. Received YYY; in original form ZZZ}
\pubyear{2018}

   \volume{{\rm in press}}

\begin{document}
\label{firstpage}
\pagerange{\pageref{firstpage}--\pageref{lastpage}}
\maketitle

\begin{abstract}
The duration of quasar accretion episodes is a key quantity for
distinguishing between models for the formation and growth of
supermassive black holes, the evolution of quasars, and their
potential feedback effects on their host galaxies.  However, this
critical timescale, often referred to as the quasar lifetime, is
still uncertain by orders of magnitude ($t_{\rm Q} \simeq 0.01\,{\rm Myr} -
1\,{\rm Gyr}$). Absorption spectra of quasars exhibiting transmission in
the \ion{He}{ii} Ly$\alpha$ forest provide a unique opportunity to
make precise measurements of the quasar lifetime.  Indeed, the size of
a quasar's \ion{He}{ii} proximity zone, the region near the quasar
itself where its own radiation dramatically alters the ionization
state of the surrounding intergalactic medium (IGM), depends
sensitively on its lifetime for $t_{\rm Q}\lesssim 30\,{\rm Myr}$,
comparable to the expected $e$-folding time-scale for SMBH growth
$t_{\rm S} =45\,{\rm Myr}$. In this study we compare the sizes of
\ion{He}{ii} proximity zones in the Hubble Space Telescope (HST) spectra of six $z \simeq 4$
quasars to theoretical models generated by post-processing
cosmological hydrodynamical simulations with a $1$D radiative transfer
algorithm. We introduce a Bayesian statistical method to infer the
lifetimes of individual quasars which allows us to fully marginalize
over the unknown ionization state of \ion{He}{ii} in the surrounding
IGM. We measure lifetimes $t_{\rm Q} = 0.63^{+0.82}_{-0.40}$~Myr and $t_{\rm Q} = 5.75^{+4.72}_{-2.74}$~Myr for two objects. For the other four quasars large redshift uncertainties
undermine our sensitivity allowing us to only place upper or lower limits.
However a joint analysis of these four systems yields a measurement
of their average lifetime of $\langle t_{\rm Q}\rangle = 1.17^{+1.77}_{-0.84}$~Myr.
We discuss our short $\sim 1\,{\rm Myr}$ inferred lifetimes in the
context of other quasar lifetime constraints and the growth of
SMBHs. 
\end{abstract}


\section{Introduction}
\label{sec:intro}

Quasars are the most luminous non-transient sources of radiation in
the Universe. It is believed that they played an important role in
the evolution of cosmic structure on all scales, for instance
dramatically changing the ionization and thermal state of the
surrounding intergalactic medium \citep{McQuinn2009, Compostella2013,
  Khrykin2016, Khrykin2017, Chardin2017, Davies2017, LaPlante2017,
  DAloisio2017}, but also possibly regulating star formation in
galaxies \citep{Springel2005a, Hopkins2006}. However, many questions
about the nature of quasars and their evolution remain unanswered. For
example, the existence of SMBHs with masses $M_{\rm BH} \simeq
10^9-10^{10}M_{\sun}$ \citep{Fan2001, Fan2004, Mortlock2011,
  Venemans2013, DeRosa2014, Wu2015, Banados2018}
already at $z \sim 6-7$ poses a
challenge for current theories of SMBH formation, requiring very
massive initial black hole seeds and accretion of matter on time-scales
comparable to the Hubble time \citep{Hopkins2009, Volonteri2010,
  Volonteri2012}. Solving this puzzle is impossible without
constraints on the characteristic time-scale over which accretion
on to SMBHs occurs.

A holy grail would thus be a direct measurement of the lifetime
$t_{\rm Q}$ of quasars at high redshift, or more precisely the
duration of their individual accretion episodes. This would shed light
on triggering and feedback mechanisms \citep{Springel2005a, Hopkins2006},
on how gas funnels to the centers of galaxies, and on the structure of
the inner accretion disc \citep{Goodman2003,Hopkins2010}. Unfortunately,
the best currently available estimates on the time-scales governing
quasar activity are uncertain by several orders of magnitude (see
\citealp{Martini2004} for a review). For instance, studies of quasar
demographics which attempt to model the quasar luminosity function
and/or quasar clustering \citep{Haiman2001,Martini2001} constrain the
total integrated time that a galaxy hosts an active quasar, known as
the quasar duty cycle $t_{\rm dc}$. These studies often come to very
different conclusions depending on the particular assumptions made in the modeling, but constraints are in the range
$t_{\rm dc} \approx 1\,{\rm Myr}-1\,{\rm Gyr}$ \citep{Yu2004, Adelberger2005,
  Croom2005, Shen2009, White2012, Conroy2013, Cen2015,
  LaPlante2016}. In particular, the very strong clustering of
quasars measured at $z \sim 4$ implies $t_{\rm dc} \sim 1$~Gyr
\citep{Shen2007, White2008}, much higher than estimates of this duty
cycle at lower-$z$, and seems to require an extremely tight
relationship between the black hole mass $M_{\rm BH}$ and the halo
mass $M_{\rm halo}$ \citep{White2008}. It is important to note that  while these
demographic studies constrain the quasar duty cycle $t_{\rm dc}$, they are 
insensitive to the duration of individual accretion episodes $t_{\rm Q}$,
which theoretical models suggest could be much shorter \citep{Ciotti2001,Novak2011}.

Previous studies employing a variety of techniques have found
tentative evidence for short episodic quasar lifetimes $t_{\rm
  Q}$. For example, recent work by
\citet{Schawinski2010,Schawinski2015} based on light travel time
arguments in active galactic nuclei (AGN) host galaxies argues for
quasar variability on time-scales of $t_{\rm Q} \simeq
0.1$~Myr. However, plausible alternative scenarios related to AGN
obscuration could explain those observations without invoking short
time-scale quasar variability, and furthermore the giant $\sim
500\,{\rm kpc}$ Ly$\alpha$ nebulae discovered around $z\sim 2$
quasars \citep{Cantalupo2014,Hennawi2015} implies quasar lifetimes of $\gtrsim 1$~Myr, at odds
with these short inferred lifetimes. \citet{Oppenheimer2018} (see also \citealp{Goncalves2008}) argued
that the high incidence of strong high-ionization absorption lines in the circumgalactic medium of galaxies
provides evidence for relic light echoes from an AGN, implying $t_{\rm Q} \lesssim 1$~Myr, but
the nature of these absorbers is a subject of intense debate and many models manage to explain them
\citep{Stern2016,Stern2018,Mathews2017,McQuinn2018}
without invoking a previously active AGN. 
Finally, the presence of high equivalent width  Ly$\alpha$ emitters (LAEs) near hyper-luminous
quasars has been attributed to quasar-powered fluorescence and invoked to argue for quasar lifetimes of $1 \lesssim t_{\rm Q} \lesssim 30$~Myr \citep{Trainor2013, Borisova2016}. However, the expected boost due to the quasar illumination is
inconsistent with the much brighter observed Ly$\alpha$ luminosities of these LAEs, strongly suggesting that these sources are powered intrinsically
and not by the nearby quasar \citep[but see][]{Adelberger2006}. 
To summarize, all of the aforementioned methods are indirect and often involve model-dependent assumptions, such that plausible alternative scenarios can be invoked to explain the observations.

On the other hand, the ionization state of the IGM in quasar environs
provides a powerful method to estimate quasar lifetimes. For example,
using the observational survey data presented in \cite{Schmidt2017},
\cite{Schmidt2018} carefully analyzed the \ion{He}{ii} Ly~$\alpha$
transverse proximity effect (TPE; \citealp{Worseck2006, Jakobsen2003,
  Schmidt2017}), which is the increased IGM \ion{He}{ii} Ly~$\alpha$ transmission in a
background quasar sightline due to the enhancement of the radiation
field from a foreground quasar. They found that the strength of the
TPE signal exhibits a degeneracy between quasar lifetime and the
solid angle of the UV emission (or equivalently the fraction of quasars that are obscured), and uncovered a
possible bimodality in quasar emission
properties whereby some quasars appear to be unobscured and relatively long
lived ($t_{\rm Q} \gtrsim 10$~Myr), whereas others are either
younger ($t_{\rm Q} \lesssim 10$~Myr) or highly
obscured. 

This degeneracy with quasar obscuration can be removed if one instead
considers the line-of-sight (LOS) proximity effect, since the
background quasar illuminates the IGM along our sightline towards
Earth by construction.  Indeed, the LOS proximity effect in the
\ion{H}{i} Ly~$\alpha$ forest at $z \simeq 2-4$
\citep{Carswell1982, Bajtlik1988} has been used to constrain $t_{\rm
  Q}$, but it provides only weak lower limits on $t_{\rm Q} \gtrsim 0.01$~Myr \citep{DallAglio2008}.
This limit results from the fact that in order to produce a detectable
LOS proximity effect the quasar must shine longer than the time it
takes the IGM to attain ionization equilibrium with the enhanced
quasar radiation, the so-called equilibration time-scale
$t_{\rm eq} \simeq \Gamma_{\rm HI}^{-1}$.  Given the measurements of the UV \ion{H}{i} background photoionization rate 
$\Gamma_{\rm HI}^{\rm bkg} \simeq 10^{-12}{\rm s^{-1}}$
\citep{Becker2013} at $z \simeq 3-5$, the \ion{H}{i} proximity effect
is, in principle, detectable, provided that $t_{\rm Q} \gtrsim 0.01$~Myr. This
argument provides the basis for the recent discovery of a population
of very young quasars ($t_{\rm Q} \lesssim 0.01-0.1$~Myr) at $z \simeq
6$ \citep{Eilers2017,Eilers2018} as inferred from small sizes of their
LOS proximity zones. It remains unclear whether a similar population
of young quasars exists at lower redshift.  It could be that they have
only been uncovered at $z \sim 6$ because the much higher Ly$\alpha$
optical depth of the surrounding IGM makes their small proximity zones
particularly conspicuous. In addition, a more precise constraint on lifetime can be obtained if one combines the analysis of high-$z$ \ion{H}{i} proximity zones with the study of damping wing signatures in the quasar spectra \citep{Davies2018}. This approach, applied to the spectra of two most distant $z \simeq 7$ quasars, provides evidence for the lifetime $t_{\rm Q} \lesssim 10$~Myr (Davies et al. in prep). 

\ctable[ caption = {Main parameters of $7$ quasars used in this study. From left to right the columns show: quasar name, quasar position, COS spectral resolution at $1450\angstrom $, signal-to-noise ratio per $0.24\angstrom $ pixel near \ion{He}{ii} Ly~$\alpha$, quasar redshift, redshift uncertainty, spectroscopic line that was used to measure the redshift, $i$-band magnitude, absolute magnitude at $1450\angstrom$, \ion{H}{i} and \ion{He}{ii} total photon production rates $Q_{\rm 1Ry}$ and $Q_{\rm 4Ry}$, measured size of the proximity zone $R_{\rm pz}$ with corresponding $1\sigma$  redshift uncertainty, and the inferred quasar lifetime $t_{\rm Q}$.},left, maxwidth=17.9cm,star = twocolumn, doinside = \scriptsize]{X c c c c c c c c c c c c c}{

\label{tab:table1}

 }{
\hline\hline  \\
Quasar & R.A. & Decl. & $R$ & S/N & $z$ & $\Delta z$ & $z$ line & $i$-mag & $M_{1450}$ & ${\rm log}_{10}Q_{\rm 1Ry}$ & ${\rm log}_{10}Q_{\rm 4Ry}$ & $R_{\rm pz} \pm \sigma \left(R_{\rm pz}\right)$ & ${\rm log}_{10} \left( t_{\rm Q}/{\rm Myr} \right)$ \\  & (degree) & (degree) & & & & ${\rm km\ s^{-1}}$ & & & & $ {\rm s^{-1}} $ & $ {\rm s^{-1}} $ & ${\rm Mpc}$ &  \\  [1.1ex]  \hline
\\ [3ex] 
HE2QS~J2311$-$1417    & $23.114546$	& $-14.17521 $ & 2300 & 4 &$3.700$ & 656 & \ion{C}{iv} & 18.11 & $-27.64$ & $57.56$ & $56.66$ & $1.94  \pm 1.72$  & $< 0.31$ \\
SDSS~J1137$+$6237     & $11.372172$ & $+62.37072$ & 2300 & 4 &$3.788$ & 656 & \ion{C}{iv} & 19.31 & $-26.46$ & $57.10$ & $56.19$   & $4.92 \pm 1.68$ & $>-0.90$ \\
HE2QS~J1630$+$0435 & $16.305634$ & $+04.35594$ & 2000 & 4 &$3.810$ & 400 & ${\rm H}~{\beta}$ & 17.51 & $-28.37$ & $57.82$ & $56.92$   & $8.43 \pm 1.02$ & $0.76^{+0.26}_{-0.28}$ \\ 
SDSS~J1614$+$4859    & $06.142681$ & $+48.59588$ & 2300 & 3 &$3.817$ & 656 & \ion{C}{iv} & 19.45 & $-26.34$ & $57.05$ & $56.14$  & $2.72 \pm 1.66$ & $<1.19$ \\ 
HE2QS~J2354$-$2033 & $23.545200$ & $-20.33207$ & 2300 & 3 &$3.774$ & 656 & \ion{C}{iv} & 18.90 &  $-26.88$ & $57.27$ & $56.37$  & $-3.65\pm 1.68$ & ---\\
SDSS~J1711$+$6052     & $17.113441$ & $+60.52403$ & 2700 & 4 &$3.835$ & 656 & \ion{C}{iv} & 19.34 & $-26.49$ & $57.10$ & $56.19$   & $2.97 \pm 1.65$ & $<1.25$\\
SDSS~J1319$+$5202    & $13.191420$ & $+52.02001$ & 2700 & 2 &$3.916$ & 400 & ${\rm H}~{\beta}$ & 17.81  & $-28.02$ & $57.73$  & $56.82$ & $3.62 \pm 0.98$ & $-0.20^{+0.36}_{-0.45}$ \\
[2ex]
\hline
}

\citet{Khrykin2016} showed that the analogous LOS \ion{He}{ii}
Ly~$\alpha$ proximity effect \citep{Hogan1997, Anderson1999, Zheng2015}
at $z \simeq 3-4$ is sensitive to quasar lifetimes on
longer and more interesting time-scales of up to $t_{\rm Q} \approx
30$~Myr, comparable to the Salpeter or $e$-folding
\citep{Salpeter1964} time-scale for SMBH growth $t_{S} = 45$~Myr.  This arises from the fact that the equilibration time-scale
for \ion{He}{ii} at these redshifts is three orders of magnitude
longer owing to the much lower \ion{He}{ii} photoionization rate
$t_{\rm eq}\approx \Gamma_{\rm HeII}^{-1} \simeq 30\, {\rm Myr}$
\citep{Khrykin2016}.  Fortunately,  significant observational effort over the last
several years 
has resulted in the discovery of large numbers of new $z\sim 3-4$ quasar
sightlines which are transparent at \ion{He}{ii} Ly$\alpha$ in the
quasar rest-frame \citep{Worseck2011, Syphers2012,
  Worseck2016}. But to date these \ion{He}{ii} proximity zones
have not yielded convincing quantitative constraints on the time-scales
governing quasar emission (but see \citealp{Zheng2015, Syphers2014}).

In this work we compare the proximity zones of six $z \simeq
3.7-3.9$ quasars from \citet{Worseck2018} to theoretical models and obtain
 the first robust quantitative constraints on quasar lifetime from
the \ion{He}{ii} Ly$\alpha$ LOS proximity effect. Our modeling builds
on \citet{Khrykin2016} where a custom 1D radiative transfer algorithm
was used to post-process outputs from a cosmological hydrodynamical
simulation, which takes into account the radiation from the quasar and
the metagalactic ionizing background which sets the ionization state of
the ambient IGM.  Whereas previous analyses of \ion{He}{ii} proximity
zones made specific assumptions about the ionization state of
\ion{He}{ii} in the ambient IGM near the quasars \citep{Syphers2014,
  Zheng2015}, \citet{Khrykin2016} argued that a degeneracy exists
between the IGM ionization state and the quasar lifetime. Here we introduce
a Bayesian method to constrain the lifetimes of \emph{individual}
quasars in this dataset, which allows us to fully marginalize over the
unknown ionization state of the \ion{He}{ii} in the IGM.

This paper is organized as follows. In Section~\ref{sec:data} we
summarize the observations and the parameters of the quasars in our dataset. 
We outline our numerical model in Section~\ref{sec:model}. In
Section~\ref{sec:sims} we describe our new Bayesian method for measuring the
quasar lifetime and \ion{He}{ii} fraction, and present the results of our
inference in Section~\ref{sec:mcmc}. We discuss our
findings in Section~\ref{sec:disc}, and conclude in
Section~\ref{sec:consl}.

Throughout this work we assume a flat $\Lambda$CDM cosmology with dimensionless
Hubble constant $h = 0.7$, $\Omega_{\rm m}=0.27$, $\Omega_{\rm
  b}=0.046$, $\sigma_{8} = 0.8$ and $n_s=0.96$ \citep{Larson2011},
and helium mass fraction $Y_{\rm He}=0.24$. All distances are quoted in
{\it proper} Mpc unless explicitly stated otherwise.

\section{Data Sample}
\label{sec:data}

\begin{figure*}
\begin{center}
\includegraphics[width=1.0\linewidth]{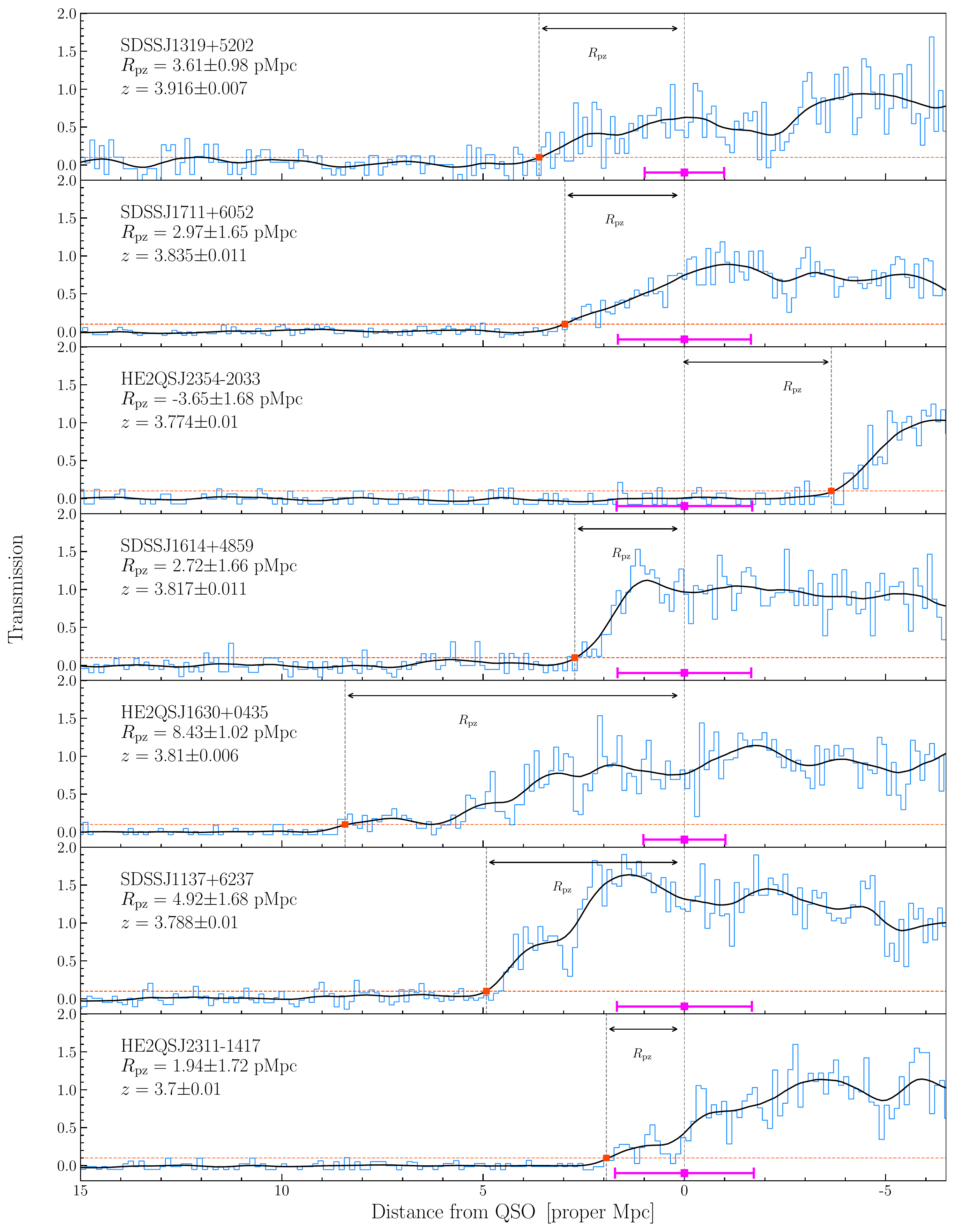}
\end{center}
\caption{\ion{He}{ii} transmission spectra of $7$ $z \simeq 4$ quasars in our data sample (see Table~\ref{tab:table1}). The blue binned lines show the HST/COS spectra ($0.24\ {\rm \angstrom/pixel}$), whilst the black lines show the $0.97$~Mpc-smoothed \ion{He}{ii} transmission, respectively. The magenta error bar illustrates the redshift uncertainty of the corresponding quasar. The values of measured proximity zone sizes are indicated by the red squares, where the smoothed transmission crosses the $10$~per cent threshold (red horizontal lines).}
\label{fig:data}
\end{figure*}

We use a sample of 7 $z\ge 3.7$ \ion{He}{ii}-transparent quasars observed with the Cosmic Origins Spectrograph
\citep[COS;][]{Green2012} on board the HST.
Table~\ref{tab:table1} summarizes their key properties. Six of them are taken from \citet{Worseck2018},
where a detailed description of the observations and the data reduction are presented. Here we summarize
the most relevant details. The spectra were taken with the \textit{HST}/COS G140L grating at different COS Detector Lifetime Positions,
resulting in somewhat different spectral resolutions $R=\lambda/\Delta\lambda=2,000$--$2,700$ at
the wavelengths of interest near \ion{He}{ii} Ly~$\alpha$ ($\lambda\simeq 1450$\,\AA).
The spectra have been rebinned to $0.24$\,\AA\,pixel$^{-1}$ yielding a sampling of 2--3 pixels per resolution element,
and the signal-to-noise ratio (S/N) per pixel near \ion{He}{ii} Ly~$\alpha$ varies from two to four. 

To this sample we add the quasar HE2QS~J2354$-$2033 that had been discovered in our
dedicated ground-based survey for UV-bright high-redshift quasars \citep{Worseck2018}.
\textit{HST}/COS G140L follow-up spectroscopy was obtained in Program 14809 (PI Worseck)
on 30 October 2016 at COS Lifetime Position 3, yielding a spectral resolution $R\simeq 2,300$
at $1450$\,\AA. A single four-orbit visit (total exposure time 11,131\,s) yielded a S/N$\simeq 3$
per $0.24$\,\AA\ pixel in the quasar continuum near \ion{He}{ii} Ly~$\alpha$. 
The individual exposures were reduced with the \textsc{CALCOS} pipeline v2.21, and then
post-processed with custom software for accurate background estimation and co-addition
in the Poisson limit, as described in \citet{Worseck2016,Worseck2018}. Custom COS detector
pulse height screening was employed to minimize the dark current while including almost all source flux.
We determined a science pulse height range 2--12 from the strong geocoronal \ion{H}{i} Ly~$\alpha$ line.
For dark current subtraction in the spectral trace we used a sample of COS dark monitoring data taken
within three months centred on the date of the science observations as in \citet{Worseck2016,Worseck2018}.

The spectra were corrected for Galactic extinction adopting the respective selective extinction $E(B-V)$ from
\citet{Schlegel1998} and the \citet{Cardelli1989} extinction curve assuming the Galactic average ratio
between total $V$ band extinction and selective extinction $R_V=3.1$. The extinction-corrected spectra
were normalized with power-law continua redward of \ion{He}{ii} Ly~$\alpha$, accounting for low-redshift
\ion{H}{i} Lyman limit breaks \citealp{Worseck2016,Worseck2018}.
Only one quasar in the sample (SDSS~J1137$+$6237) shows signs of \ion{He}{ii} Ly$\alpha$ emission
that we do not incorporate in the continuum model. Its neglect, as well as other (percent-level) continuum errors
do not affect our measurements of the proximity zone size.

In order to estimate the size of the proximity zone ($R_{\rm pz}$) for each quasar spectra we follow previous conventions used in the studies of \ion{H}{i} proximity zones \citep{Fan2006, Bolton2007, Lidz2007, Carilli2010, Eilers2017} and define $R_{\rm pz}$ to be the location in the spectrum where the smoothed \ion{He}{ii} transmission profile drops below $10$~per cent for the first time. \citet{Fan2006} used a Gaussian filter with ${\rm FWHM} = 20 $\angstrom \ (in the observed frame) to smooth the spectra.  This smoothing scale corresponds to a velocity interval $\Delta v \sim 700\ {\rm km\ s^{-1}}$ or a distance interval $\Delta R = 0.97$ proper Mpc at $z \simeq 6$. We adopt the same smoothing scale $\Delta R = 0.97$~proper Mpc at redshifts of the quasars in the data sample, and apply a Gaussian filter to all \ion{He}{ii} Ly$\alpha$ transmission profiles. Because the smoothing length is larger than the spectral resolution and includes several pixels, the measured proximity zone size does only weakly depend on the S/N of the spectra, and we do not include these effects into our numerical calculations below. Fig.~\ref{fig:data} illustrates the \ion{He}{ii} transmission profiles of each quasar in our sample. The smoothed \ion{He}{ii} transmission is indicated by the black lines. The magenta error bars indicate the corresponding redshift uncertainty of each quasar.

The dominant source of uncertainty in our proximity zone measurements
results from quasar redshift errors. It is well known that the primary
broad rest-frame UV/optical emission lines that are accessible in the
optical/near-IR for $z > 3.5$ quasars have line centres which can
differ from systemic by as much as $\sim 3000\,{\rm km\,s^{-1}}$ due
to outflowing/inflowing material in the so-called broad line regions
of quasars \citep{Gaskell1982, Tytler1992, VandenBerk2001, Richards2002,
  Shen2016}. To robustly determine systemic redshifts and associated
errors for the quasars in our sample, we adopt an approach similar to
that described in \citet{Shen2016}. The idea is to use a training sample
of quasars for which the broad lines in question are present, as well
as other features which are known to be good tracers of the systemic
frame. For example, \citet{Shen2016} used spectra of $z < 1$ quasars
taken from Sloan Digital Sky Survey Reverberation Mapping (SDSS-RM)
project to determine the relationship between the broad ${\rm H}~\beta $ $\lambda 4861$
emission line redshift and systemic redshift determined via
\ion{Ca}{ii} K $\lambda 3934$ absorption lines arising from the quasar host galaxy.  For
cases where the only strong broad far-UV lines like \ion{C}{iv} $\lambda 1549$ are
available, as is the case at $z > 3.5$ if only optical spectra are
available, then one can similarly use a training set to
calibrate the relationship between \ion{C}{iv} redshifts and those
determined from the lower ionization (but still broad) \ion{Mg}{ii}
$\lambda 2798$ emission line, which is known to be an excellent tracer of the
systemic frame \citep{Shen2016}.

Of the seven quasars in our sample, two objects SDSS~J1319$+$5202 and
HE2QS~J1630$+$0435 have high-quality near-IR spectra which enable a
redshift measurement from the rest-frame optical ${\rm H}~\beta$ lines which
are redshifted into the $K$-band (Worseck et al in prep.). For the other five quasars in our
sample which lack high-quality near-IR spectra, we determine redshifts
from the \ion{C}{iv} emission line, which is the strongest line
detected redward of \ion{H}{i} Ly~$\alpha$. The broad ${\rm H}~\beta$ and \ion{C}{iv}
emission lines were centred using the algorithm described in
\citet{Hennawi2006}, which is robust against spectral noise, line
asymmetries, and associated absorption features (which can affect
\ion{C}{iv}). For the two objects with ${\rm H}~\beta$ redshifts, we
transformed into the systemic frame using the average blueshift of
$-109\,{\rm km\,s^{-1}}$ and assigned them a redshift error of
$400\,{\rm km\,s^{-1}}$, following the values derived by
\citet{Shen2016}. For the other five with \ion{C}{iv} the procedure is
slightly more complicated. It is well known that the blueshift of the \ion{C}{iv} line
is luminosity dependent, which is known as the Baldwin effect
\citep{Baldwin1977}. Thus to obtain systemic redshifts from \ion{C}{iv}
we follow \citet{Shen2016} and use a training set to fit for the 
luminosity dependent blueshift between \ion{C}{iv} and \ion{Mg}{ii}
redshifts.
To briefly summarize how we calibrated the
luminosity dependent \ion{C}{iv} blueshift, 
we used a sample of 2504 quasars from the Baryon Oscillation Spectroscopic Survey \citep[BOSS;][]{BOSS2018}
with sufficient wavelength coverage and ${\rm S\slash N}$
ratio that redshifts from both \ion{C}{iv} and \ion{Mg}{ii} can be obtained. These quasars have monochromatic
luminosities at $1450\angstrom$ in the range $\log_{10} L_{1450}\simeq 43.5-47$, where  $\log_{10} L_{1450}$ is determined from the $i$-band apparent magnitudes of the quasars and a composite quasar spectrum following the
approach described in Appendix A of \citet{Hennawi2006}. To quantify the luminosity dependent velocity shift
we bin these data in luminosity with a bin size of $\Delta \log_{10} L = 0.25$ and compute the mean shift in each
bin after sigma clipping outliers. The error on the mean is determined from the standard deviation
and divided by $\sqrt{N}$, where $N$ is the number of points considered, i.e. those that survived
sigma clipping. Following \citet{Shen2016} we fit a simple linear relation for the dependence of
these velocity shifts on $\log L_{1450}$
\begin{equation}
  v = a + b {\rm log}_{10} L_{1450} - {\rm log}_{10} L_{1450,0},
\end{equation}
where $\log_{10} L_{1450,0} = 45$. This procedure yields $a = -192.4\,{\rm
  km\,s^{-1}}$ and $b =-599.6\,{\rm km\,s^{-1}}$, which is in
reasonable agreement with the independent result obtained by
\citet{Shen2016} for a distinct training set, and different line
centering algorithms. Using this fit, we then transform every
\ion{C}{iv} redshift in this training set into the systemic frame (here
defined by the \ion{Mg}{ii} redshift). An error on this procedure can
be estimated by considering the distribution of these estimated
\ion{Mg}{ii} redshifts about their true values. We find that this
distribution is well described by a Gaussian with a mean $-79\,{\rm
  km\,s^{-1}}$ and standard deviation $656\,{\rm km\,s^{-1}}$, and we
use the latter as the error on our \ion{C}{iv} redshifts. Note that for
these \ion{C}{iv} redshifts, we simply assume that \ion{Mg}{ii} frame
perfectly traces systemic. Neglecting the small differences between
the \ion{Mg}{ii} frame is a valid assumption given 
the results from
\citet{Shen2016}, who found that the distribution of \ion{Mg}{ii} about
systemic (defined by the [\ion{O}{ii}] emission line) is well described
by a Gaussian with mean shift $-20\,{\rm km\,s^{-1}}$ and standard
deviation $200\,{\rm km\,s^{-1}}$ --- both of which are much smaller than our
inferred error budget of $656\,{\rm km\,s^{-1}}$ arising from the imperfect
correlation between \ion{C}{iv}-\ion{Mg}{ii} and redshifts. 

Unfortunately, in case of quasar HE2QS~J2354$-$2033, its redshift appears to be significantly underestimated.  From the extent of the \ion{He}{ii} proximity zone we conclude that the \ion{C}{iv} emission line of this quasar, that was used to measure the systemic redshift, has a very large blueshift. This is clearly seen in Fig.~\ref{fig:data}, where the resulting \ion{He}{ii} proximity zone has a hugely negative size ($R_{\rm pz} = -3.65\pm 1.68$~pMpc). Regrettably, there are no other strong lines in the spectrum of HE2QS~J2354$-$2033 redward of \ion{H}{i} Ly~$\alpha$ that can be used to accurately determine the redshift of this quasar. Therefore, we exclude the quasar HE2QS~J2354$-$2033 from the subsequent analysis after the initial inspection of the \ion{He}{ii} proximity zone, because it cannot provide reasonable constraint on the lifetime. 

The redshifts, associated redshift errors, and the emission line used to infer the redshift are given in
Table~\ref{tab:table1}. The full details of our procedure, as well as information about the
near-IR spectra and the data used for each quasar redshift 
are provided in \citet{Worseck2018}.

\begin{figure*}
\centering
 \includegraphics[width=1.0\linewidth]{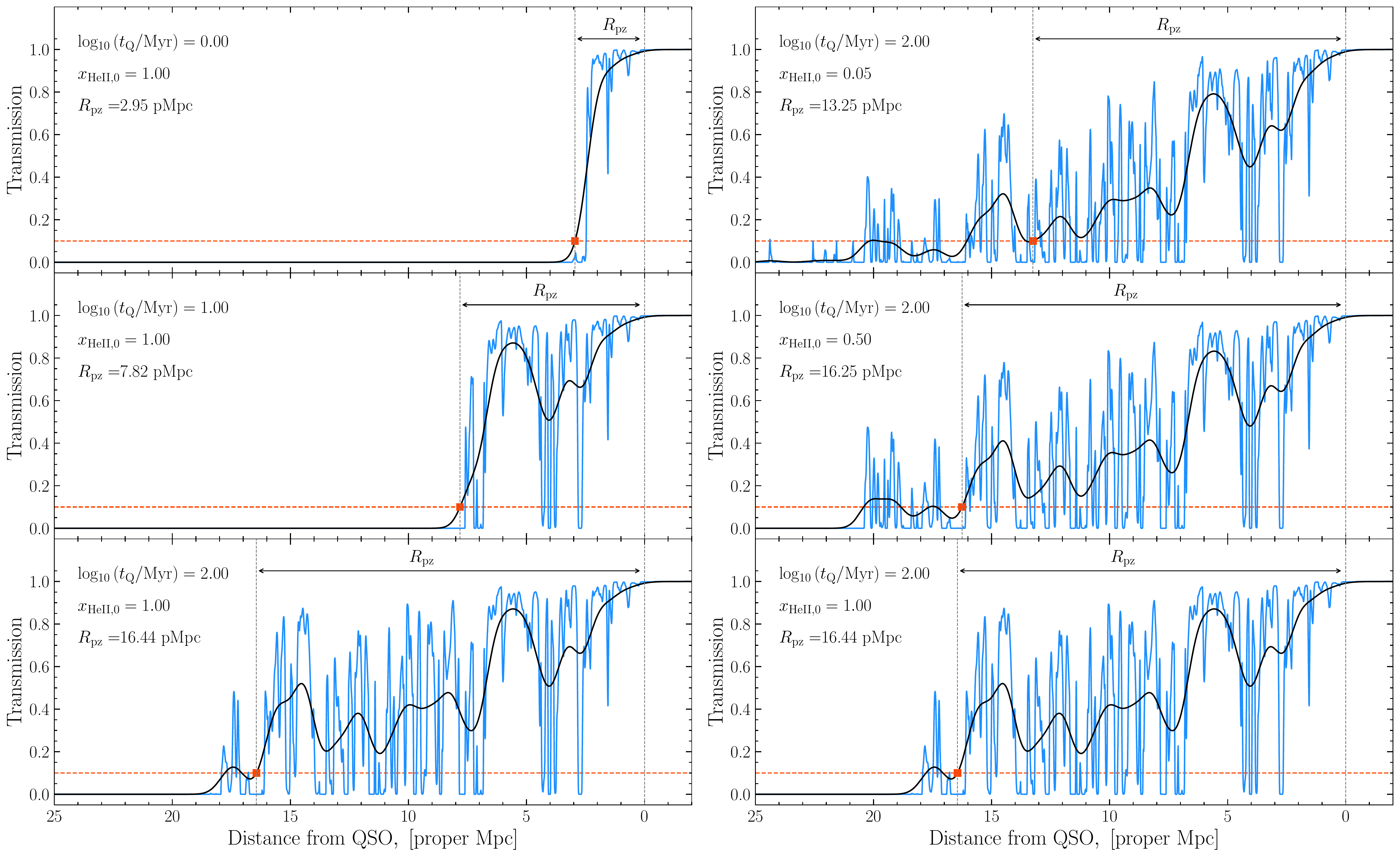}
 \caption{Examples of \ion{He}{ii} transmission profiles in our radiative transfer simulations of quasar SDSS~J1319+5202 at redshift $z = 3.916$ and $Q_{\rm 4Ry} = 10^{56.82}\ {\rm s^{-1}}$ (see Table~\ref{tab:table1}). The left columns show \ion{He}{ii} transmission spectra (blue) and smoothed \ion{He}{ii} transmission profiles (black) in models with initial \ion{He}{ii} fraction $x_{\rm HeII,0} = 1.00$ and three different quasar lifetimes, whereas the right column shows the case of a varying initial \ion{He}{ii} fraction, but a fixed quasar lifetime ${\rm log}_{10} \left( t_{\rm Q} / {\rm Myr} \right) = 2.0$. The {\it red} horizontal lines indicate the $10$~per cent transmission threshold, and the corresponding sizes of \ion{He}{ii} proximity zones (marked by red squares) are indicated by the black arrows.}
 \label{fig:heII_trans}
\end{figure*}

\section{Modeling \ion{He}{II} Ly~$\alpha$ Proximity Zones}
\label{sec:model}
 
Our numerical model consists of hydrodynamical simulations and a $1$D post-processing radiative transfer algorithm for transport of the ionizing radiation from the quasar through the IGM. In this section we provide the most important details of our model and refer the reader to the full description given in \citet{Khrykin2016, Khrykin2017}

We use the Gadget-3 code \citep{Springel2005b} with simulation box size of $25h^{-1}$~comoving
Mpc on a side, containing $2 \times 512^3$ particles. Using periodic boundary conditions we extract $1000$ density, velocity, and temperature fields (skewers) drawn in random directions around the most massive haloes ($M > 5\times 10^{11}M_{\sun}$) in the outputs of hydrodynamical simulations at $z = 3.7$ and $z = 3.9$.  The resulting skewers have a total length of $160$~{\it comoving} Mpc with a pixel scale $ {\rm d} r = 0.01$~{\it comoving} Mpc (${\rm d} v = 1.0\ {\rm km\ s}^{-1}$). 

Extracted skewers are used in our $1$D post-processing radiative transfer algorithm based on the ${\rm C}^2$-Ray code \citep{Mellema2006}, which calculates the evolution of the abundances of ${\rm e}^{-}$, \ion{H}{i},
\ion{He}{ii}, and the gas temperature \citep{Khrykin2016, Khrykin2017}.  Assuming there
is no evolution of cosmic structure between the redshifts of the
simulation  outputs and that of the corresponding quasars, we simply
rescale the gas density of the skewers by a factor $\left( 1 + z
\right)^3$ to account for cosmological density evolution, where $z$ is
the redshift of the quasar that we are simulating (see
Table~\ref{tab:table1}). 

  \begin{figure*}
\centering
 \includegraphics[width=1.0\linewidth]{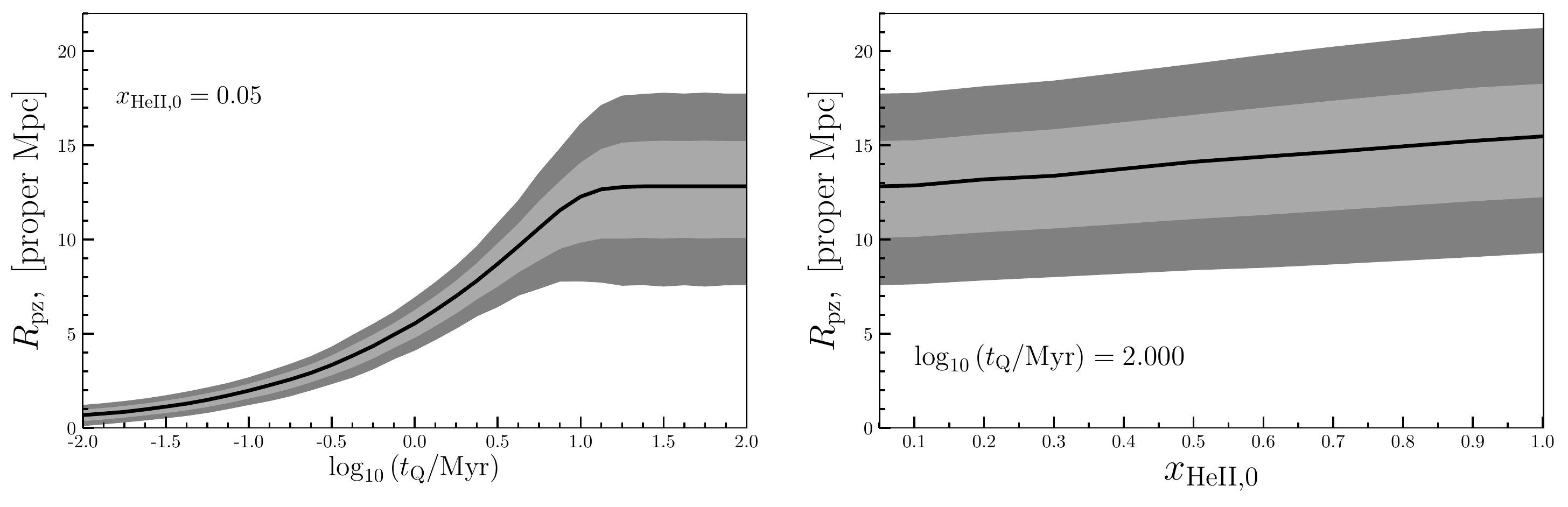}
 \caption{Dependence of the \ion{He}{ii} proximity zone size $R_{\rm pz}$ on quasar lifetime $t_{\rm Q}$ (left panel, with fixed $x_{\rm HeII,0} = 0.05$) and initial \ion{He}{ii} fraction $x_{\rm HeII,0}$ (right panel with fixed ${\rm log}_{10} \left( t_{\rm Q} / {\rm Myr} \right) = 2.0$) in radiative transfer simulations for quasar SDSS~J1319$+$5202. The light and dark grey shaded areas show $1\sigma$ and $2\sigma$ standard deviations of $R_{\rm pz}$, respectively, whereas the solid black line illustrates the median values.}
 \label{fig:R_pz_tQ_xHeII}
\end{figure*}

There are several important parameters which govern quasar
\ion{He}{ii} proximity zones: the quasar systemic redshift, the photon production rates
$Q_{\rm 1 Ry}$ and $Q_{\rm 4Ry}$ above ${\rm 1Ry}$ and ${\rm 4Ry}$,
respectively, the quasar lifetime $t_{\rm Q}$, and the \ion{He}{ii}
ionizing background, which sets the value of the initial \ion{He}{ii}
fraction $x_{\rm HeII,0}$, which prevailed in the IGM before the
quasar turned on \citep{Khrykin2016, Khrykin2017}\footnote{In what follows we
quote the values of initial \ion{He}{ii} fraction that correspond to
the adopted values of the \ion{He}{ii} background in the radiative
transfer simulations}.  For each quasar in our data sample we create a
custom set of radiative transfer models based on these parameters.
Using their observed $i$-band magnitudes, 
we compute $Q_{\rm 1 Ry}$ and
$Q_{\rm 4Ry}$ for each quasar in our sample (see
Table~\ref{tab:table1}) according to the procedure outlined in
\citet[see also Section~\ref{sec:disc}]{Hennawi2006}. These values, together with the systemic redshift $z$ are fixed for all
radiative transfer models of each individual quasar in our data
sample. On the other hand, we explore different combinations of the parameters
\{$t_{\rm Q}, x_{\rm HeII,0} $\}. We consider logarithmically
spaced quasar lifetime values in the range ${\rm log}_{10} \left( t_{\rm
  Q} / {\rm Myr} \right) = \left[ -2.0, 2.0\right]$, with $\Delta
{\rm log}_{10} \left( t_{\rm Q} / {\rm Myr} \right) = 0.125$, whereas
the initial \ion{He}{ii} fraction can have one of the following values
$x_{\rm HeII,0} = \left[ 0.05, 0.10, 0.20, 0.30, 0.50, 0.60, 0.70,
  0.90, 1.00 \right]$.  This results in a grid of $297$ radiative transfer
models, with $1000$ \ion{He}{ii} Ly~$\alpha$ transmission spectra per model, 
for each quasar in our data sample.

\section{Estimating the Quasar Lifetime and Initial \ion{He}{II} Fraction}
\label{sec:sims}

\subsection{Mock \ion{He}{II} Spectra and the Size of the Proximity Zone}
\label{sec:mocks}

Following the same procedure applied to the observational data,
we smooth our simulated 
\ion{He}{ii} Ly~$\alpha$ transmission spectra with a Gaussian with
width of $0.97$~proper Mpc (see Section~\ref{sec:data}).
Examples of mock \ion{He}{ii} transmission spectra smoothed in this
way are shown by the black curves in Fig.~\ref{fig:heII_trans},
whereas the blue curves show the original full-resolution mock spectra that
results from our modeling procedure. The proximity zone sizes are marked by the
red squares.
It is apparent from the left side panels of Fig.~\ref{fig:heII_trans} that, as expected, the size of the proximity zone increases for longer quasar lifetimes. This dependence of the proximity zone size on quasar lifetime exists because the IGM responds to the changes in the radiation field and attains a new ionization equilibrium state on a finite equilibration time-scale, which is  $t_{\rm eq} \simeq 1 / \Gamma_{\rm bkg}^{\rm HeII} \approx 30$~Myr for \ion{He}{ii} at $z \simeq 3-4$ \citep{Khrykin2016}. The left panel of Fig.~\ref{fig:R_pz_tQ_xHeII} illustrates the dependence of $R_{\rm pz}$ on quasar lifetime determined from
our mock spectra with different lifetimes. 
It is apparent that for short quasar lifetimes ($t_{\rm Q} \leq t_{\rm eq}$) the
median proximity zone size  $R_{\rm pz}$ grows with lifetime, until $t_{\rm Q}$ becomes larger than the equilibration time ($t_{\rm Q} \geq t_{\rm eq}$),  at
which point the median $R_{\rm pz}$ saturates and stops growing \citep{Khrykin2016}.

A much weaker increasing trend can be seen in the right panels of Fig.~\ref{fig:heII_trans} where the $R_{\rm pz}$ is larger for higher initial \ion{He}{ii} fraction. This trend may seem counter-intuitive upon examination of
the full-resolution transmission profiles (blue curves), which exhibit
more transmission at larger distances from the quasar for lower values
of initial \ion{He}{ii} fraction. This is because the shape of the
transmission profile depends on both the initial \ion{He}{ii} fraction
which is set by the ionizing background, as well as the  amount of
photoelectric heating of the IGM resulting from the reionization of
\ion{He}{ii}, which is known as the thermal proximity effect
\citep[see also \citealp{Bolton2009,Meiksin2010}]{Khrykin2017}.
The overall increase in transmission at larger
radii from the quasar in models with low $x_{\rm HeII,0}$ is caused by
the higher \ion{He}{ii} background as compared to models with $x_{\rm
  HeII,0} > 0.50$.  \citet{Khrykin2016} showed that the effect of \ion{He}{ii}
background, which sets the value of the initial \ion{He}{ii} fraction,
on \ion{He}{ii} proximity zone becomes prominent at distances where
the quasar photoionization rate $\Gamma_{\rm HeII}^{\rm QSO}$ is no longer
the dominant contribution to the total photoionization rate, i.e., $\Gamma_{\rm
  HeII}^{\rm QSO} \lesssim \Gamma_{\rm HeII}^{\rm bkg}$. However, the
standard definition of the proximity zone size that we use, which is
when the smoothed transmission crosses $10$~per cent, is insensitive
to the information about $\Gamma_{\rm HeII}^{\rm bkg}$ encoded in the \ion{He}{ii}
transmission at the outskirts of the proximity zones. This is because at larger
distances where  $\Gamma_{\rm HeII}^{\rm QSO}$ becomes comparable to the background the
transmission is much lower than $10$~per cent, which is visible in the
right panels of Fig.~\ref{fig:heII_trans}
(a similar argument explains why the \ion{H}{i} Ly$\alpha$ proximity zones of $z\sim 6$
quasars are not good probes of $x_{\rm HI}$ and reionization \citep{Eilers2017}). 
On the other hand, the photoionization of \ion{He}{ii}
by the quasar causes significant heating of the IGM gas. The
more singly ionized helium present in the surrounding IGM, the more
photoelectric heating results from the absorption of hard photons.
Because of the $\tau_{\rm Ly\alpha} \propto T^{-0.7}$ temperature dependence
of the \ion{He}{ii} Ly~$\alpha$ optical depth, this heating boosts the
transmission in the proximity zone for
higher values of $x_{\rm HeII,0}$
\citep{Khrykin2017}. Therefore, the smoothed transmission profiles
cross the $10$~per cent threshold at larger radii in case of higher $x_{\rm
  HeII,0}$ values. 
  
  \begin{figure*}
\centering
 \includegraphics[width=1.0\linewidth]{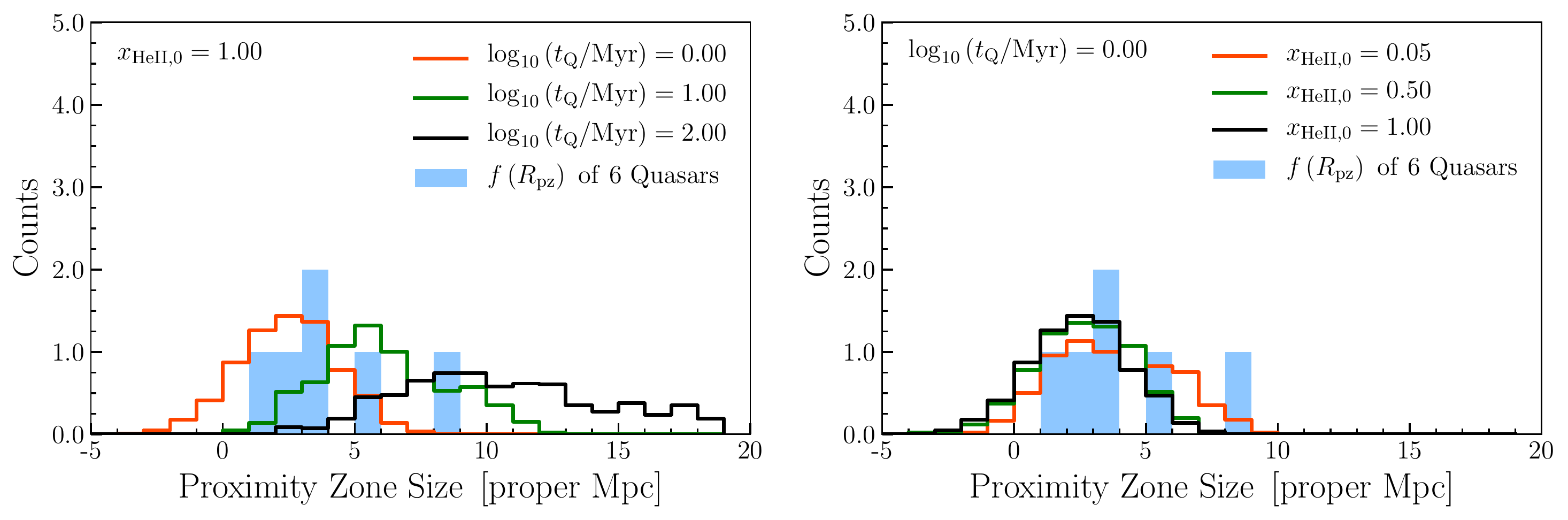}
 \caption{Example of $f\left( R_{\rm pz}\right)$ distributions in radiative transfer simulations. Similar to Fig.~\ref{fig:heII_trans}, The left column illustrates the $R_{\rm pz}$ distributions in models with varying quasar lifetime and $x_{\rm HeII,0} = 1.00$, whereas the right column shows the case of fixed quasar lifetime at  ${\rm log}_{10} \left( t_{\rm Q} / {\rm Myr} \right) = 0.00$ and different values of the initial \ion{He}{ii} fraction. The blue histogram in both panels illustrates the proximity zone sizes of quasars in our data sample. }
 \label{fig:heII_trans_hist}
\end{figure*}

\begin{figure*}
\centering
 \includegraphics[width=1.0\linewidth]{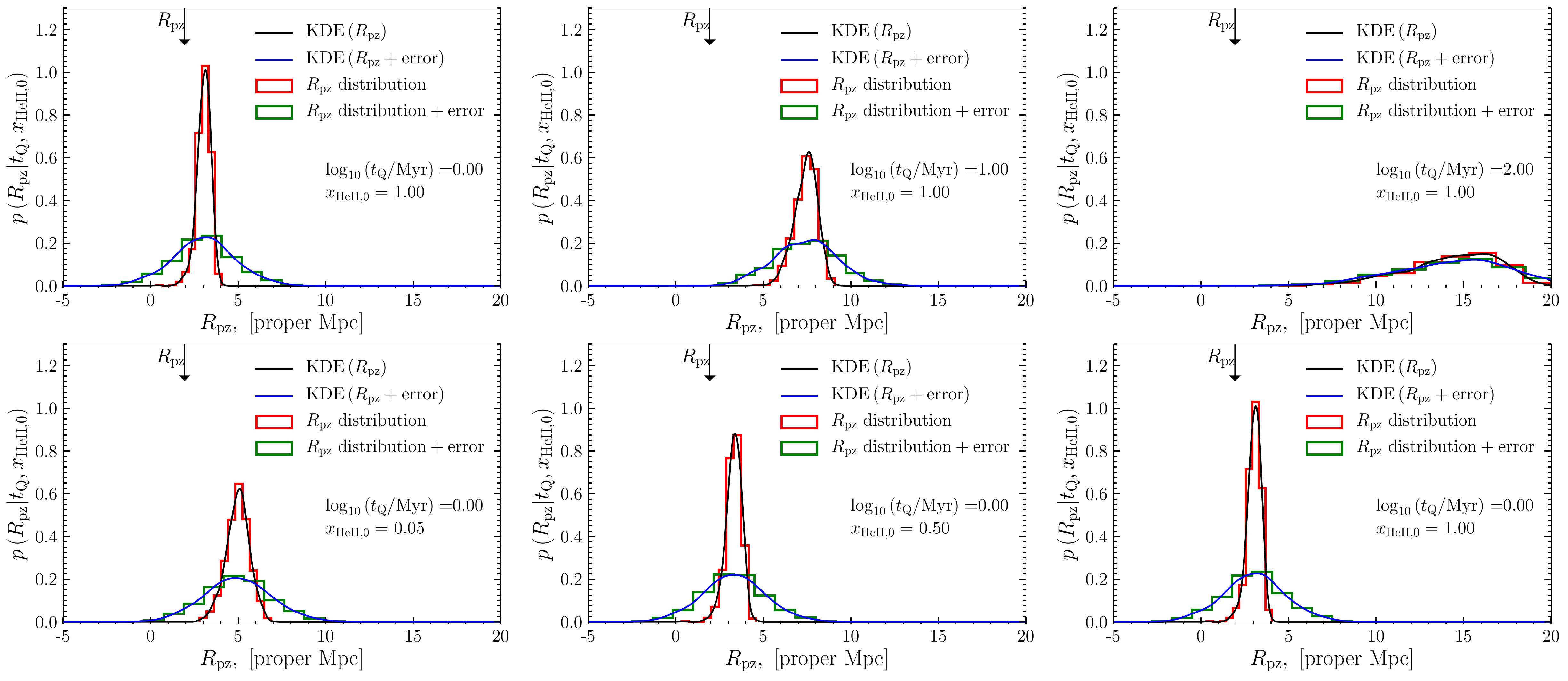}
 \caption{Example kernel density estimation on several distributions of simulated $R_{\rm pz}$. The top row illustrates the results for several models with varying quasar lifetimes, and a fixed value of the initial \ion{He}{ii} fraction, whereas the bottom row shows similar results, but now for a fixed value of quasar lifetime and different $x_{\rm HeII,0}$. The red and green histograms in each panel illustrate the simulated $R_{\rm pz}$ distributions before and after adding redshift errors. The black and blue curves are the respective smooth KDE functions of $R_{\rm pz}$. The arrow in each panel indicates the value of $R_{\rm pz}$ for quasar HE2QS~J2311$-$1417 (see Table~\ref{tab:table1}).}
 \label{fig:kde}
\end{figure*}

The competition between the increased transmission in the ambient IGM for small $x_{\rm HeII,0}$,
the $10$~per cent transmission threshold being too high to be probe $\Gamma_{\rm HeII}^{\rm bkg}$, 
and the increased transmission due to the thermal proximity effect for higher $x_{\rm HeII,0}$ results in an overall weak dependence of proximity zone size with initial \ion{He}{ii} fraction, illustrated in the right panel of Fig.~\ref{fig:R_pz_tQ_xHeII}.

\subsection{Distribution of the Proximity Zone Sizes}
\label{sec:rpz}

In general, density fluctuations in the IGM give rise to 
a broad distribution of proximity zone sizes $R_{\rm pz}$ for
a given set of model parameters \citep{Khrykin2016}, as illustrated in Fig.~\ref{fig:R_pz_tQ_xHeII}.
Furthermore, the uncertainty in quasar redshifts
also adds a significant amount of scatter. This is apparent from
Table~\ref{tab:table1} where it is seen (second to last column) that our
largest redshift errors, which are those derived from the \ion{C}{iv} emission line,
result in an uncertainty $\sigma(R_{\rm pz}) \simeq 1.7\,{\rm Mpc}$, comparable
to the smallest measured proximity zone sizes. Notwithstanding these uncertainties, a comparison of the  $R_{\rm pz}$ distributions from our radiative transfer models to
our data sample can still yield constraints on the quasar lifetime. This is readily
apparent from inspection of Fig.~\ref{fig:heII_trans_hist}, where we show the observed
distribution of $R_{\rm pz}$ along with the predicted distributions from our simulations
for different combinations of model parameters
$t_{\rm Q}$ and $x_{\rm HeII}$.
Similar to Fig.~\ref{fig:heII_trans},  the left panel of
Fig.~\ref{fig:heII_trans_hist} shows the distribution of $R_{\rm
  pz}$ for models with a fixed initial \ion{He}{ii} fraction ($x_{\rm
  HeII,0} = 1.00$), and three different values of quasar lifetime 
whereas the right panel shows results at fixed quasar lifetime ${\rm
  log}_{10} \left( t_{\rm Q} / {\rm Myr} \right) = 0.00$ and three
different values of $x_{\rm HeII,0}$. Each model histogram of
simulated \ion{He}{ii} proximity zone sizes is derived from $600$
mock $R_{\rm pz}$ measurements - where we take $100$ $R_{\rm pz}$ 
from the model of each of the six quasars for the corresponding ${\rm
  log}_{10} \left( t_{\rm Q} / {\rm Myr} \right)$ and $x_{\rm HeII,0}$
values. These model histograms are then normalized to six, which is
the total number of quasars in our data sample.
We incorporate redshift errors into our simulated distributions of $R_{\rm pz}$ using the uncertainties on $\sigma(R_{\rm pz})$ reported in Table~\ref{tab:table1}. Specifically, we randomly draw a Gaussian distributed redshift
error using the $\sigma(R_{\rm pz})$ specific to each quasar and add these to the  $100$ 
simulated $R_{\rm pz}$ values for each quasar and each combination of model parameters.

Inspection of the left panel of Fig.~\ref{fig:heII_trans_hist} already reveals that
lifetimes around $t_{\rm Q}\sim 1\,{\rm Myr}$ appear to be preferred
by the data. Whereas the right panel clearly indicates the high degree of overlap between the simulated histograms for different values of $x_{\rm HeII}$. This overlap is a direct result of the weak dependence of $R_{\rm pz}$ on initial \ion{He}{ii} fraction discussed in the previous section (see Fig.~\ref{fig:R_pz_tQ_xHeII}). This weak sensitivity combined with redshift uncertainties suggests it will be challenging to infer the $x_{\rm HeII,0}$ for these quasars. We come back to this question in Section~\ref{sec:mcmc}.

\begin{figure*}
\centering
 \includegraphics[width=1.0\linewidth]{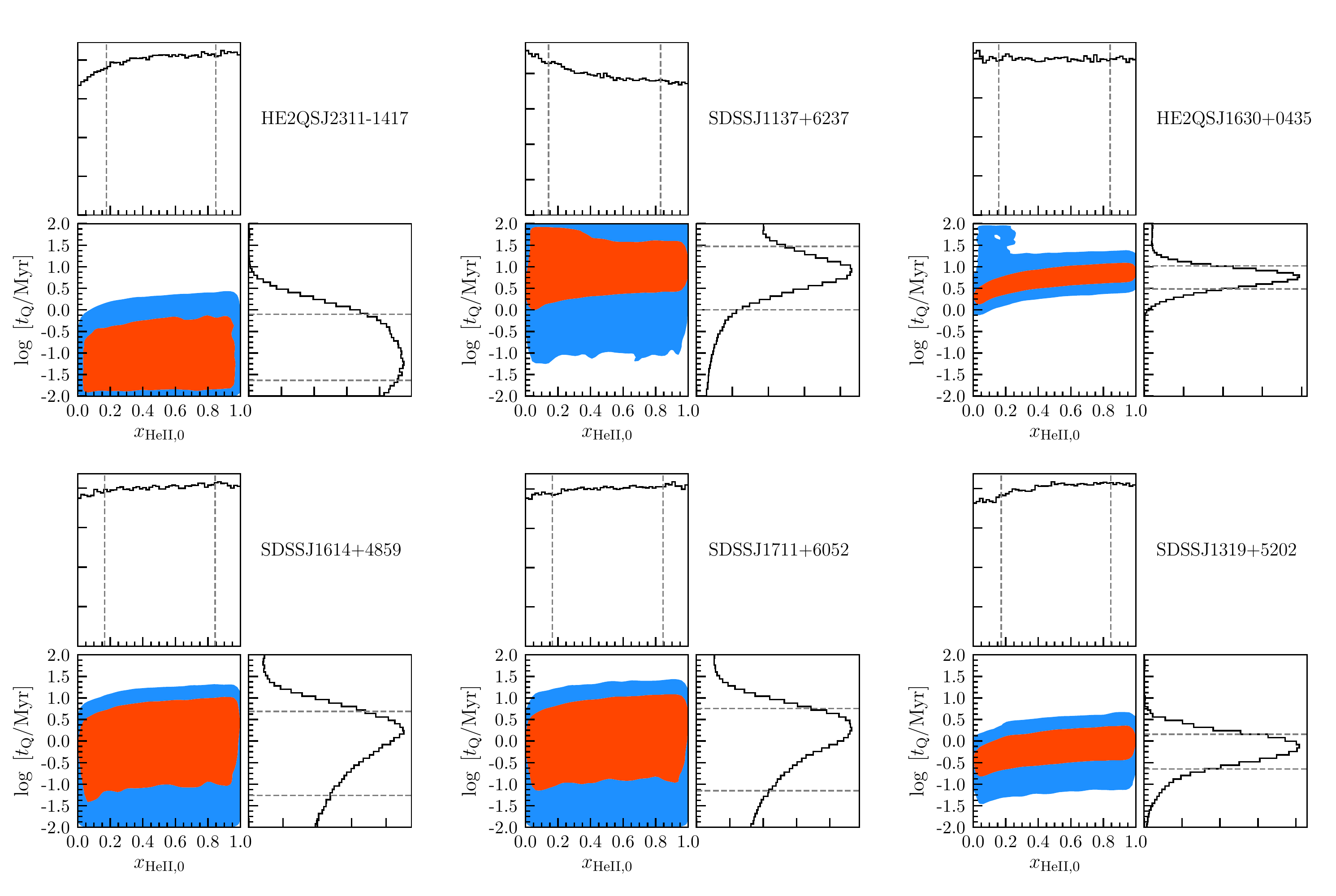}
 \caption{Constraints on the quasar lifetime and the initial \ion{He}{ii} fraction from the MCMC analysis of the 6 quasar in the data sample. The $95$ ({\it red}) and $68$~per cent ({\it blue}) confidence levels from the MCMC calculations are shown. The histograms illustrate the corresponding marginalized posterior probability distributions of each parameter.}
 \label{fig:MCMC}
\end{figure*}

\subsection{Bayesian Inference of Model Parameters}
\label{sec:like}

In order to estimate the lifetime and \ion{He}{ii} fraction
for each quasar in our data sample
we introduce a Bayesian likelihood for the observed proximity zone
size $R_{\rm pz}$ given these model parameters
\begin{equation}
\mathscr{L}\left( R_{\rm pz} | t_{\rm Q}, x_{\rm HeII,0} \right)  = p\left( R_{\rm pz} | t_{\rm Q}, x_{\rm HeII,0} \right),
\label{eqn:likelihood1} 
\end{equation}
where $p\left( R_{\rm pz} | t_{\rm Q}, x_{\rm HeII,0} \right)$ is the
probability density function (PDF) for $R_{\rm pz}$ determined from
our radiative transfer simulations for a given value of $t_{\rm Q}$
and $x_{\rm HeII,0}$ (we omit $z$, $Q_{\rm 1Ry}$, $Q_{\rm 4Ry}$ in the
notation for simplicity because these parameters are fixed for all
models of any given quasar). Note that given the finite number of
skewers used in each model, $p\left( R_{\rm pz} | t_{\rm Q}, x_{\rm
  HeII,0} \right)$ must be estimated from a discrete number of
samples. To this end we use kernel density estimation (KDE) to
estimate the PDF from the 1000 simulated $R_{\rm pz}$ values for each
model, which results in a smooth continuous approximate function
$p\left( R_{\rm pz} | t_{\rm Q}, x_{\rm HeII,0}
\right)$. Fig.~\ref{fig:kde} shows an example of this KDE procedure
and the resulting PDFs for a set of radiative transfer models of
quasar HE2QS~J2311-1417, where the histogram of the $R_{\rm pz}$ values
is shown in red, and the corresponding KDE models are plotted as black
lines.

As stated previously, redshift uncertainty constitutes a significant
source of error, which would alter the outcome of the parameter inference if neglected.
We therefore model the impact of these redshift uncertainties on the $R_{\rm pz}$
PDF by adding random Gaussian distributed redshifts errors with
standard deviation $\sigma(R_{\rm pz})$ (see Table~\ref{tab:table1})
to the simulated $R_{\rm pz}$ of each quasar, and perform the KDE on these noisy
values.  The results are illustrated in Fig.~\ref{fig:kde}, where
$R_{\rm pz}$ distributions with redshift uncertainty ($\sigma(R_{\rm pz}) = 1.72~{\rm pMpc}$ for HE2QSJ2311-1417, see Table~\ref{tab:table1}) included 
are shown by green histograms, and the corresponding KDEs are illustrated by the
blue lines. It is clear that incorporating redshift uncertainties into our
$R_{\rm pz}$ distributions makes them broader and reduces the
discriminating power of each individual proximity zone size measurement. 

In order to calculate the likelihood of the data for any combination
of model parameters, we can then simply evaluate the corresponding KDE
PDF at the observed value of $R_{\rm pz}$ (see Table~\ref{tab:table1})
for the quasar in question. This procedure results in $297$
determinations of the likelihood at each location \{$ t_{\rm Q},
x_{\rm HeII,0}$\} on our model grid for each quasar. We use bivariate
spline interpolation to compute $\mathscr{L}\left( R_{\rm pz} | t_{\rm
  Q}, x_{\rm HeII,0} \right)$ for any combination of \{$ t_{\rm Q},
x_{\rm HeII,0}$\} between the model grid points in our parameter
space.  

Armed with the above likelihood, we can now conduct Bayesian inference
of model parameters for each quasar using Markov Chain Monte Carlo
(MCMC).  Given our lack of knowledge about the \ion{He}{ii}
background, which sets the initial \ion{He}{ii} fraction, at the
redshifts considered in this work, we choose a flat linear prior on
$x_{\rm HeII,0}$ from $x_{\rm HeII,0} = 0.00$ to $x_{\rm HeII,0} =
1.00$. On the other hand, we set a flat logarithmic prior on ${\rm
  log}_{10}\left(t_{\rm Q} \slash {\rm Myr} \right)$ from $-2.0$ to
$2.0$. The lower value of ${\rm log}_{10}\left( t_{\rm Q} \slash {\rm
  Myr} \right) =-2.0$ is motivated by the ubiquitous presence of the
LOS proximity effect in the \ion{H}{i} Ly$\alpha$ forest (but see
\citealp{Eilers2017}), which implies lifetimes in excess of $t_{\rm Q}
\gtrsim 1\slash \Gamma_{\rm HI} \simeq 0.01\,{\rm Myr}$ for the vast
majority of quasars. The upper value of ${\rm log}_{10}\left( t_{\rm
  Q} \slash {\rm Myr} \right)=2.0$ is chosen as it lies in the upper
range of lifetime estimates in the literature based on both quasar
duty cycle and black hole growth arguments \citep[see][for a
  review]{Martini2004}. Furthermore, for $t_{\rm Q}$ in excess of
$100\,{\rm Myr}$ several of the assumptions that we are making in the
modeling, like our neglect of cooling, and our post-processing
approach which implicitly assumes that cosmic structure is fixed over
the time-scales that the quasar radiation alters its environment,
begin to break down. 

 We will describe the results of our MCMC
inference in the next section, where we will also see that our results
are not hugely sensitive to this choice of priors.

\section{Results}
\label{sec:mcmc}

Given the likelihood of our data given the model parameters
in equation~(\ref{eqn:likelihood1}), and our interpolation procedure
which allows us to evaluate this likelihood at any point in our
parameter space \{$
t_{\rm Q}, x_{\rm HeII,0}$\}, we now proceed to sample this likelihood
with MCMC to determine the posterior distribution of the
model parameters
for each quasar in the data sample. To this end we use the publicly
available Python MCMC software {\it emcee} \citet{Foreman2013}, which
implements an affine invariant MCMC ensemble sampling algorithm \citep{Goodman2010}.
The results of the MCMC sampling of the posterior distribution of each quasar in our data sample are shown in Fig.~\ref{fig:MCMC}, where the contours illustrate the $95$ (blue) and $68$~per cent (red) confidence intervals, respectively. Marginalized posterior probability distributions for each parameter ${\rm log}_{10} \left( t_{\rm Q} / {\rm Myr}\right)$ and $x_{\rm HeII,0}$ are also shown by the histograms. There are several noticeable results and trends that we
now discuss. 

\subsection{Constraints on the Initial \ion{He}{II} Fraction}
\label{sec:mcmc_xHeII}

We begin with the constraints on the initial \ion{He}{ii} fraction. As was previously stated in Section~\ref{sec:rpz}, the large degree of overlap between the model $R_{\rm pz}$ distributions for different values of $x_{\rm HeII,0}$ and 
fixed $t_{\rm Q}$ (see Fig.~\ref{fig:heII_trans_hist}), which ultimately
results from the very weak dependence of $R_{\rm pz}$ on $x_{\rm HeII,0}$ shown in Fig.~\ref{fig:R_pz_tQ_xHeII},  suggests that it would be difficult to infer
$x_{\rm HeII,0}$ with the current dataset. This is indeed the case -- the broad flat posterior distributions for $x_{\rm HeII,0}$ in Fig.~\ref{fig:MCMC}, which hardly differ from our assumed flat prior, indicate that the data is not very informative and the initial \ion{He}{ii} fraction in the ambient IGM surrounding our seven quasars is
virtually unconstrained.

\subsection{Constraints on the Quasar Lifetime}
\label{sec:mcmc_logtQ}

Fig.~\ref{fig:posterior_tQ} shows the one dimensional posterior
probability distributions for all seven quasars in our sample,
marginalized over the initial \ion{He}{ii} fraction. It is clear upon
inspection of these posteriors that in some cases we are able to
measure the quasar lifetime, whereas in others we can only set upper
limits. Clearly this distinction
depends upon the strength of the peak in the posterior probability.
In order to distinguish between measurements and limits, we use
the following criterion.  If the maximum value of the marginalized posterior probability distribution is at least four times larger than the larger of the two posterior probability values at the edges
of the ${\rm log}_{10} \left(t_{\rm Q} / {\rm Myr}\right)$ parameter grid, then we classify it
as a measurement. In this case we quote the $50$th percentile of the posterior distribution as the
measured value, whereas the $16$th and $84$th percentiles are quoted as our uncertainties.
On the other hand, for the flatter posterior probability distributions which do not satisfy the
above criteria, we report an upper limit on the quasar lifetime.  We choose to quote 
the $95$th percentile value as our upper limit (effectively $2\sigma$) on the lifetime. 

One issue with the upper limits as we define them is that they clearly
depend on the range of simulated ${\rm log}_{10} \left( t_{\rm Q} /
{\rm Myr}\right)$ values, i.e., on our choice of a flat logarithmic
prior on the quasar lifetime extending from ${\rm log}_{10} \left(
t_{\rm Q} / {\rm Myr}\right) = -2.0$ to ${\rm log}_{10} \left(
t_{\rm Q} / {\rm Myr}\right) = 2.0$. However, as was discussed in
Section~\ref{sec:like}, the lower limit of our prior is physically
motivated observations of the \ion{H}{i} LOS proximity effect.  The
upper limit of our prior is determined by limitations of our modeling
procedure, but we see that the proximity zones are all so small that
the resulting posteriors are all very small at ${\rm log}_{10} \left(
t_{\rm Q} / {\rm Myr}\right) = 2.0$, and thus it does not influence
our results. 

 The results of our lifetime inference for all quasars in
our data sample are summarized in Table~\ref{tab:table1}, which we
discuss further in what follows.
 
\subsubsection{SDSS~J1319$+$5202}
\label{sec:5202}

The red histogram in Fig.~\ref{fig:posterior_tQ} illustrates the
marginalized lifetime posterior probability distribution for the quasar SDSS~J1319$+$5202.
We infer the lifetime of ${\rm log}_{10} \left( t_{\rm Q} / {\rm Myr}\right) = -0.20^{+0.36}_{-0.45}$, indicating that this quasar is relatively young. It is also clear from the shape of the posterior that for this case of highly constraining
data we are totally insensitive to our choice of priors.
The reason we are able to constrain $t_{\rm Q}$ so tightly for
SDSS~J1319$+$5202 is that it is the second most luminous object in our sample,
which also has the smallest redshift uncertainty $\sigma(R_{\rm pz}) =
0.98$~Mpc as inferred from its H$\beta$ emission line redshift (see
Table~\ref{tab:table1}). 

Proximity zone sizes $R_{\rm pz}$
increase with luminosity \citep{Khrykin2016}, implying that
the relative error on $R_{\rm pz}$ should be smallest for the brightest
sources. This appears to be reflected in SDSS~J1319$+$5202 which has
the third largest proximity zone size of $R_{\rm pz} = 3.62 \pm 0.98$~Mpc. 
In contrast with the green histograms (blue curves) in Fig.~\ref{fig:kde} for
HE2QS~J2311$-$1417 which has a $89\%$ relative error on  $R_{\rm pz}$, the $27\%$ relative
error on $R_{\rm pz}$ for SDSS~J1319$+$5202 implies its  $R_{\rm pz}$
PDFs are significantly less broadened by the redshift uncertainty. As a result, the likelihood values, obtained by evaluating the respective KDE (see Section~\ref{sec:like}), are less similar for different $t_{\rm Q}$ models, resulting in higher
lifetime precision.

\subsubsection{HE2QS~J1630$+$0435}
\label{sec:0435}
\begin{figure}
\centering
 \includegraphics[width=1.0\linewidth]{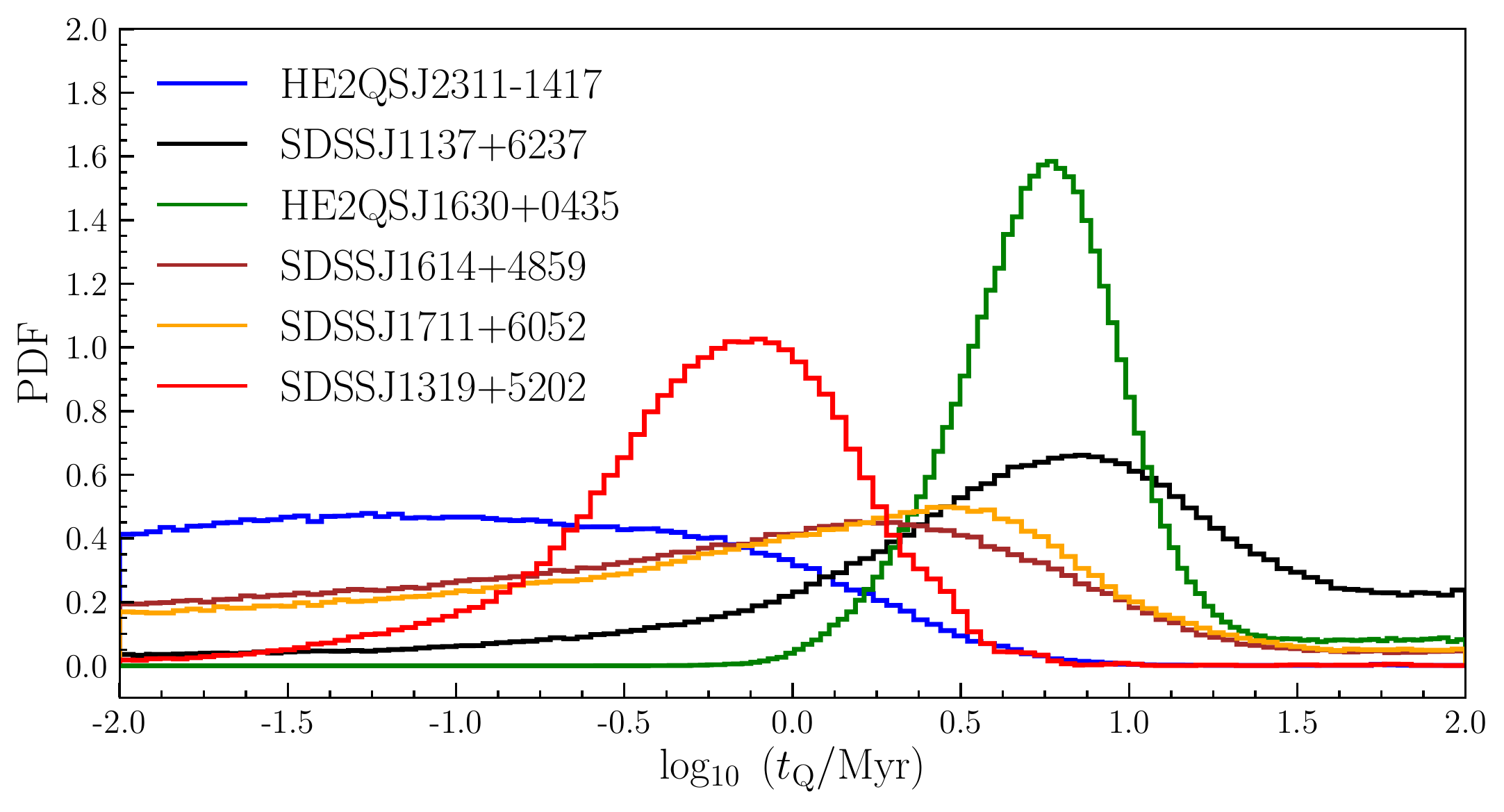}
  \caption{One-dimensional posterior probability distributions of ${\rm log}_{10} \left( t_{\rm Q}/ {\rm Myr}\right)$ from MCMC calculations marginalized over the initial \ion{He}{ii} fraction. Each histogram represents the results for a quasar in our data sample.}
\label{fig:posterior_tQ}
\end{figure}

It is apparent from the green histogram in Fig.~\ref{fig:posterior_tQ} that the best lifetime constraint we obtain is
for the most luminous quasar in the sample HE2QS~J1630$+$0435, which has $R_{\rm pz} = 8.43 \pm 1.02$~Mpc. 
 From this distribution we deduce the lifetime to be ${\rm
  log}_{10} \left( t_{\rm Q} / {\rm Myr}\right) =
0.76^{+0.26}_{-0.28}$.  Similar to SDSS~J1319$+$5202, the redshift
error for this quasar translates into a small uncertainty of $\sigma(R_{\rm pz}) = 1.02$~Mpc owing to
the H$\beta$ emission line redshift. This constitutes a $\simeq 12$~per cent relative
error on $R_{\rm pz}$, which allows us to constrain the lifetime with good precision. 

\subsubsection{SDSS~J1137$+$6237}
\label{6237}

The black histogram in Fig.~\ref{fig:posterior_tQ} shows the marginalized posterior probability distribution for the lifetime of quasar SDSS~J1137$+$6237. This quasar has the second biggest proximity zone ($R_{\rm pz} = 4.92 \pm 1.68$~Mpc) in our sample. Taking into account its relatively low luminosity, it is possible that it has a long lifetime. However, while this quasar has only $\approx 34\%$ relative error on the proximity zone size (comparable to $\approx 27\%$ error for SDSS~J1319$+$5202), which, in principle, should provide good constraint on quasar lifetime, the resulting posterior distribution looks different than in case of SDSS~J1319$+$5202 and HE2QS~J1630$+$0435. Namely, the long tail of similar high posterior probabilities at ${\rm log}_{10}\left( t_{\rm Q} / {\rm Myr} \right) > 1.25$, that are only $\approx 2.5$ times lower than the peak of the distribution, does not allow us to make a clear measurement of quasar lifetime, according to the criterion defined in Section~\ref{sec:mcmc_logtQ}.

This decreased constraining power at high $t_{\rm Q}$ values arises
from the fact that sensitivity of $R_{\rm pz}$ measurements to quasar
lifetime is limited by the value of the equilibration timescale
$t_{\rm eq}$ (see Section~\ref{sec:mocks} for more details), which is
$t_{\rm eq} \simeq 25$~Myr at $z \simeq 4$. For lifetimes
longer than the equilibration timescale $t_{\rm Q} \gtrsim t_{\rm
  eq}$, proximity zone size $R_{\rm pz}$ saturates and no longer
increases with increasing $t_{\rm Q}$ as shown in the left panel of
Fig.~\ref{fig:R_pz_tQ_xHeII}. Consequently, the $R_{\rm pz}$ PDFs for
$t_{\rm Q} \gtrsim t_{\rm eq}$ are comparable, as are the estimated
likelihoods of models with $t_{\rm Q} \gtrsim t_{\rm eq}$. Therefore,
our inference cannot distinguish between these models, which results
in a tail in the posterior distribution, clearly seen in
Fig.~\ref{fig:posterior_tQ} for SDSS~J1137$+$6237. Given the shape of
its posterior, we can only quote a $95\%$ lower limit on its
lifetime. To this end we calculate the 5th percentile of the posterior
distribution for quasar SDSS~J1137$+$6237, which yields ${\rm
  log}_{10}\left( t_{\rm Q} / {\rm Myr} \right) > -0.90$.

\subsubsection{The Remaining Quasars}
\label{sec:rest_qso}

It is apparent from Table~\ref{tab:table1} that for the remaining
quasars the uncertainties $\sigma(R_{\rm pz})$ arising from redshift
error are a much larger fraction of their proximity zone sizes. The resulting
broadening of the model $R_{\rm pz}$ PDFs (see Fig.~\ref{fig:kde})
results in weaker constraints as illustrated by the flatter less peaked posterior
distributions of these quasars in
Fig.~\ref{fig:posterior_tQ}. As a result, following our definition
of a measurement versus an upper limit, we can only provide $95$~per cent upper
limits on $t_{\rm Q}$ for these quasars. Although in one case
this limit is particularly strong, namely the small proximity zone
$R_{\rm pz} = 1.94 \pm 1.72$~Mpc of HE2QS~J2311$-$1417, combined with its
relatively high luminosity yields a $95$~per cent upper limit of
$\log_{10} \left(t_{\rm Q} / {\rm Myr}\right) < 0.31$, strongly ruling
out long lifetimes $>10$~Myr.
 
\begin{figure}
\centering
 \includegraphics[width=1.0\linewidth]{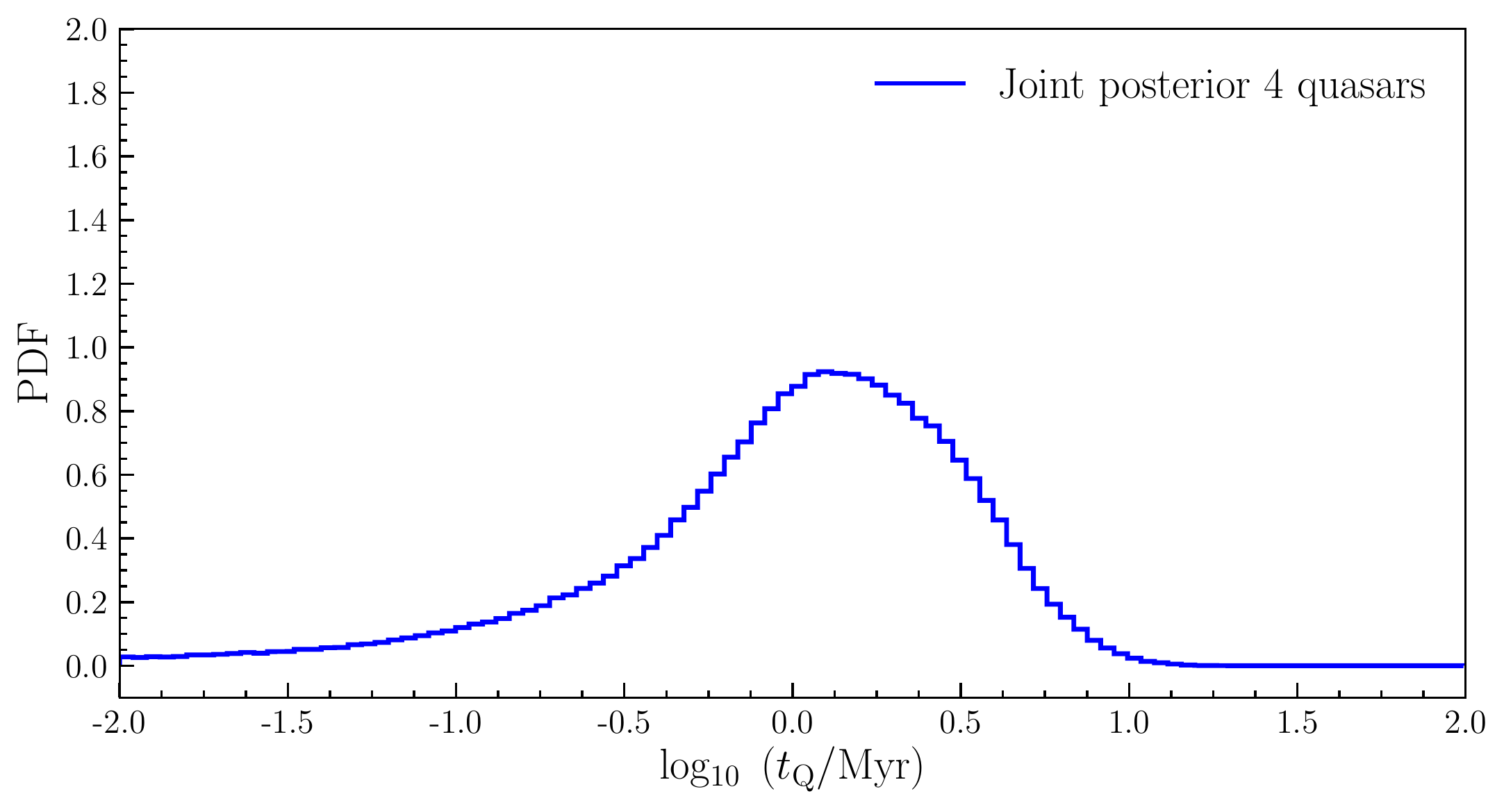}
 \caption{One-dimensional posterior probability distribution of ${\rm log}_{10} \left( t_{\rm Q}/ {\rm Myr}\right)$ from joint MCMC calculations of $4$ quasars, marginalized over the initial \ion{He}{ii} fraction. }
 \label{fig:MCMC_joint}
\end{figure}

Given the weaker constraints for the four quasars for which we only
quote limits (HE2QS~J2311$-$1417, SDSS~J1137$+$6237,
SDSS~J1614$+$4859, \& SDSS~J1711$+$6052, see Table~\ref{tab:table1}),
we decided to run a joint analysis on all of them in order to
constrain an average or effective quasar lifetime for this
subsample. Even if lifetime has some dependence on quasar luminosity, the fact
that the photon production rates for these four objects span a small
dynamic range $0.5$~dex (see Table~\ref{tab:table1}) makes this a
reasonable exercise. To conduct this joint analysis we simply multiply the
likelihoods of models corresponding to the same combinations of
\{$t_{\rm Q}, x_{\rm HeII,0}$\} for each individual quasar in the
subsample
\begin{equation}
\mathscr{L}^{\rm joint}\left( R_{\rm pz,i} | t_{\rm Q}, x_{\rm HeII,0} \right)  = \prod_{i=1}^{4}  p_i\left( R_{\rm pz} | t_{\rm Q}, x_{\rm HeII,0} \right). 
\label{eqn:likelihood2} 
\end{equation}
We then sample this joint likelihood with MCMC.  The result is
illustrated in Fig.~\ref{fig:MCMC_joint}, where we show the posterior
for $t_{\rm Q}$ marginalized over $x_{\rm HeII}$.  We find that the
effective lifetime for this subsample is $\langle {\rm log}_{10}
\left( t_{\rm Q} / {\rm Myr}\right) \rangle =
0.07^{+0.40}_{-0.55}$. This result appears consistent with the
measured lifetimes for the two quasars SDSS~J1319$+$5202 and
HE2QS~J1630$+$0435 (see Table~\ref{tab:table1}). Note that
interpreting the effective lifetime deduced by this 'stacking'
approach can be a delicate issue if there is a broad distribution of
the lifetimes \citep[see the discussion in Section 5.1
  of][]{Khrykin2016}, but in general it provides a reasonable
representation of the average lifetime of a sample.

\section{Discussion}
\label{sec:disc}

In this section we discuss systematics and several assumptions that went into our lifetime estimates and how they are relevant for placing robust constraints on quasar lifetime and initial \ion{He}{ii} fraction in the IGM. 

\subsection{Uncertainties in the Photon Production Rate}
\label{sec:qso_alpha_lum}

We have demonstrated how the radial extent of the \ion{He}{ii} proximity zone constrains the quasar lifetime. However, the size of the \ion{He}{ii} proximity zone $R_{\rm pz}$ also depends on the quasar spectral energy distribution (SED, see \citealp{Khrykin2016}), which we have assumed to be determined perfectly by the apparent magnitudes of the quasars and an assumed spectral slope. In what follows we discuss how the uncertainties in quasar SED change the constraining power of our method.

As discussed in \citet{Khrykin2016}, we approximate the quasar SED by a power-law, with specific photon production rate $N_{\nu}$ at frequencies $\nu$ above $4{\rm Ry}$ given by
\begin{equation}
N_{\nu}  =  N_{\rm 4Ry} \left( \frac{\nu}{\nu_{ \rm 4Ry}} \right)^{-\left(1 + \alpha_{\rm 4Ry \rightarrow \infty} \right)}. 
\label{eqn:Nnu}
\end{equation}
It is apparent from equation~(\ref{eqn:Nnu}) that the quasar SED is determined by: 1) the slope $\alpha_{\rm 4Ry \rightarrow \infty}$ of the SED at energies above $4$~Ry, and 2) the amplitude of the SED, given by the specific photon production rate at $4$~Ry, $N_{\rm 4 Ry}$.

First, the slope $\alpha_{\rm 4Ry \rightarrow \infty}$ determines the number of hard photons, and might affect the thermal and ionization states of IGM in quasar proximity, modifying the resulting \ion{He}{ii} transmission profile and value of the $R_{\rm pz}$ (see Section~\ref{sec:mocks}). Unfortunately, this slope is currently not determined, and in this work we assumed that the same power law slope $\alpha_{\rm 1Ry \rightarrow \infty} = 1.5$ governs the quasar spectrum for photon energies above $1$~Ry. However, in \citet{Khrykin2016} we investigated the impact of the variation in the slope $\alpha_{\rm 4Ry \rightarrow \infty}$ on the resulting \ion{He}{ii} transmission profiles. In order to capture the effect of the slope we fixed the quasar specific luminosity $N_{\rm 4Ry}$, and then freely varied the spectral slop in range $\alpha_{\rm 4Ry \rightarrow \infty} = 1.1-2.0$. We found that due to weak dependence of the quasar \ion{He}{ii} photoionization rate on $\alpha_{\rm 4Ry \rightarrow \infty}$, i.e. $\Gamma_{\rm HeII}^{\rm QSO} \propto \left( \alpha_{\rm 4Ry \rightarrow \infty} + 3\right)^{-1}$, the variations in the slope have at most a $10$~per cent effect on $\Gamma_{\rm HeII}^{\rm QSO}$. Consequently, the resulting \ion{He}{ii} transmission profiles are essentially unaffected by these variations (see fig.~12 in \citealp{Khrykin2016}). For this reason, we conclude that neither the size of the \ion{He}{ii} proximity zone $R_{\rm pz}$, nor the results of our MCMC inference will change significantly.

On the other hand, variations in the amplitude of the SED, $N_{\rm 4Ry}$, might have a profound effect on inferred values of $R_{\rm pz}$ and the results of the MCMC inference \citep{Khrykin2016}. We estimate $N_{\rm 4Ry}$ for each quasar in the data sample by scaling the observable quasar specific luminosity $N_{\rm 1Ry}$ at the \ion{H}{i} ionization threshold of $1$~Ry (determined by the corresponding $i$-band magnitudes and redshifts of the quasars in Table~\ref{tab:table1}) to the \ion{He}{ii} ionization threshold with a spectral slope $\alpha_{\rm 1Ry \rightarrow 4Ry}$ \citep{Hennawi2006, Khrykin2016}. Therefore, the variations of the amplitude $N_{\rm 4Ry}$ are inflicted by the uncertainty in $\alpha_{\rm 1Ry \rightarrow 4Ry}$. Recently, there were several studies that reported although consistent, but slightly different slopes in the extreme-ultraviolet (EUV) at $\lambda \leq 912$\AA~  \citep{Scott2004, Shull2012, Lusso2015}. Our fiducial value $\alpha_{\rm 1Ry \rightarrow 4Ry} = 1.5$ is slightly harder, but nevertheless consistent with the recent result from \citet{Lusso2015}, who found $\alpha_{\rm EUV} = 1.7 \pm 0.61$ (see also \citealp{Stevans2014}).
In what follows we explore how the uncertainty in $\alpha_{\rm 1Ry \rightarrow 4Ry}$ changes the amplitude of quasar SED and how this affects the results of our inference. To that end, we adopt a much softer slope $\alpha_{\rm 1Ry \rightarrow 4Ry} = 2.0$, consistent with new measurement by \citet{Lusso2018}.
We create a new set of radiative transfer models (similar to the discussion in Section~\ref{sec:model}) for quasar SDSS~J1319$+$5202, and perform the same analysis as in Section~\ref{sec:sims}. Finally, we run the MCMC inference on the resulting $R_{\rm pz}$ PDFs for $\alpha_{\rm 1Ry \rightarrow 4Ry} = 2.0$ case.

\begin{figure}
\centering
 \includegraphics[width=1.0\linewidth]{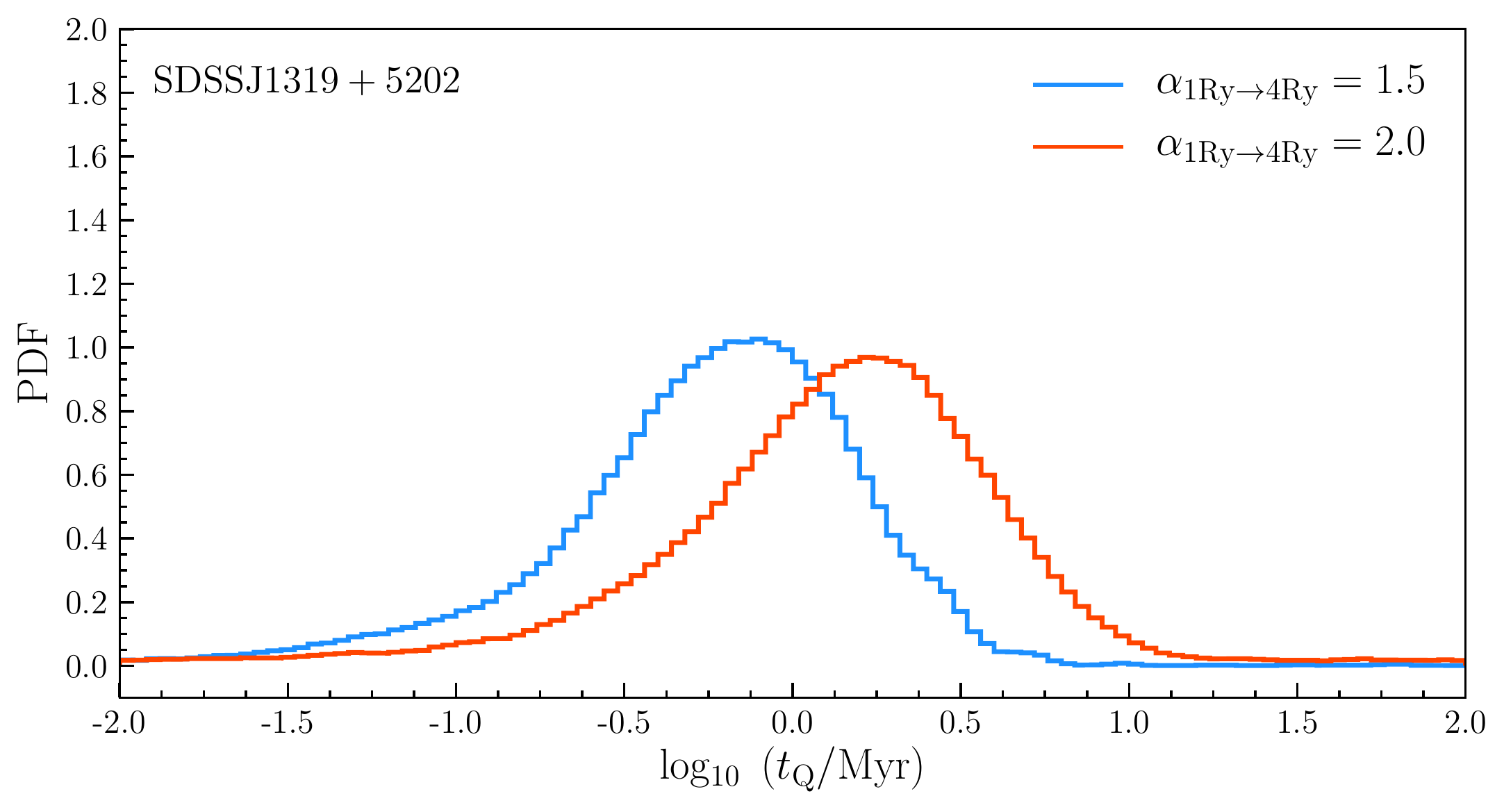}
 \caption{One-dimensional posterior probability distributions of ${\rm log}_{10} \left( t_{\rm Q}/ {\rm Myr}\right)$ from MCMC calculations of quasar SDSS~J1319$+$5202 marginalized over the initial \ion{He}{ii} fraction. The blue histogram shows the results of the inference performed on radiative transfer simulations with quasar spectral slope between $1$~Ry and $4$~Ry set to our fiducial value $\alpha_{\rm 1Ry \rightarrow 4Ry} = 1.5$ (same as in Fig.~\ref{fig:posterior_tQ}), whereas the red histogram illustrates the results for modified slope $\alpha_{\rm 1Ry \rightarrow 4Ry} = 2.0$  (see discussion in the text for more details).}
 \label{fig:MCMC_a2}
\end{figure}

Fig.~\ref{fig:MCMC_a2} shows the resulting posterior probability distribution from the MCMC inference for
quasar SDSS~J1319$+$5202 with $\alpha_{\rm 1Ry
  \rightarrow 4Ry} = 2.0$, compared to our previous findings (see Fig.~\ref{fig:posterior_tQ}) with $\alpha_{\rm 1Ry
  \rightarrow 4Ry} = 1.5$ .  We infer the lifetime ${\rm log}_{10}
\left( t_{\rm Q} / {\rm Myr}\right) = 0.17^{+0.39}_{-0.49}$. It is
apparent that the deduced lifetime of SDSS~J1319$+$5202 is $\approx
0.4$~dex longer than what we found in case of a harder spectral slope
$\alpha_{\rm 1Ry \rightarrow 4Ry} = 1.5$ (see Section~\ref{sec:mcmc_logtQ}). 

This result is expected given that the increase in $\alpha_{\rm 1Ry \rightarrow 4Ry}$ reduced the amplitude of quasar SED $N_{\rm 4Ry}$ by $0.4$~dex ($Q_{\rm 4Ry}^{\rm new} = 10^{56.4}{\rm s^{-1}}$) in our radiative transfer models. Recall that we define the size of the proximity zone as the location where the smoothed transmission profile drops below $10$~per cent for the first time. The transmission, on the other hand, is proportional to the evolution of the \ion{He}{ii} fraction, governed by equation \citep{Khrykin2016}
\begin{equation}
x_{\rm HeII} = x_{\rm HeII,eq} + \left( x_{\rm HeII,0} - x_{\rm HeII,eq} \right) e^{-\frac{t_{\rm Q}}{t_{\rm eq}}},
\end{equation}
where $x_{\rm HeII,0} \approx n_{\rm e}\alpha_{\rm HeII}\slash \Gamma_{\rm HeII}^{\rm bkg}$ and $x_{\rm HeII,eq} \approx n_{\rm e}\alpha_{\rm HeII}\slash \left( \Gamma_{\rm HeII}^{\rm QSO} + \Gamma_{\rm HeII}^{\rm bkg}\right)$ are the initial and equilibrium \ion{He}{ii} fractions. Here $n_{\rm e}$ and $\alpha_{\rm HeII}$ are the electron density and recombination coefficient, respectively. The characteristic equilibration time-scale $t_{\rm eq}$ is given by
\begin{equation}
t_{\rm eq} = \left( n_{\rm e} \alpha_{\rm HeII} + \Gamma_{\rm HeII}^{\rm QSO} + \Gamma_{\rm HeII}^{\rm bkg} \right)^{-1} \propto N_{\rm 4Ry}^{-1}
\label{eqn:t_eq}
\end{equation}
Therefore, according to equation~(\ref{eqn:t_eq}), decreasing $N_{\rm 4Ry}$ is equivalent to changing the time variable, and results in a re-scaling of our lifetime constraints by the same factor that $N_{\rm 4Ry}$ has changed. 

\subsection{Comparison to Other Constraints on Quasar Lifetime}
\label{sec:young_qso}

Our study of six $z \simeq 4$ quasars suggests lifetimes of order
$t_{\rm Q} \simeq 1$~Myr. In what follows we discuss these results
in the context of recent quasar lifetime measurements at $z\sim 6$,
and their implications for the evolution of SMBH in the high-$z$
Universe.

\citet{Eilers2017} analyzed the \ion{H}{i} proximity effect in the spectra of $34$ $z \simeq 6$ quasars and reported the discovery of three objects with exceptionally small \ion{H}{i} proximity zones that imply lifetimes $t_{\rm Q} \lesssim 0.1$~Myr, with one particular quasar shining for only $t_{\rm Q} \lesssim 0.01$~Myr \citep{Eilers2018}. These findings essentially constrain the fraction of young ($t_{\rm Q} \lesssim 0.01$~Myr) quasars to be $3$~per cent. 
Moreover, Davies et al. in prep reported the upper limit on the total duty cycle of the most distant ULAS~J1342$+$0928 quasar ($z = 7.54$; \citealt{Banados2018}) to be $t_{\rm Q} \lesssim 5.4$~Myr, based on the IGM damping wing analysis.
Consider a simple light bulb light curve model, in which the quasar is assumed to emit at constant luminosity for its entire lifetime $t_{\rm Q}$. If one randomly samples such light curves with $t_{\rm Q} \simeq 1$~Myr as suggested by our measurements, the probability of finding quasars that are as young as $t_{\rm Q} \lesssim 0.01$~Myr is $1$~per cent, which is consistent with the $\approx 3$~per cent young fraction determined
by \citet{Eilers2017} given their statistical error on one object, 
suggesting that our results are in broad agreement with their discovery of young quasars. 

However, such short lifetimes $t_{\rm Q} \sim 1$~Myr may be
problematic given the constraints from quasar clustering and current
theories about how SMBHs grow. Indeed, if one assumes a light bulb
model for the quasar light curve, then under this assumption the
lifetime $t_{\rm Q}$ and the duty cycle $t_{\rm dc}$ are equivalent,
and thus we measure the duty cycle as well.  In this case our findings
appear to be at odds with the high values of $t_{\rm dc}$ implied by
the strong clustering quasars at $z \simeq 4$ measured by
\citet{Shen2007}. For instance, \citet{White2008} modeled the
\citet{Shen2007} clustering strength and found $t_{\rm dc} \simeq
1$~Gyr, but argued that these long duty cycles are not unexpected
from the standpoint of black hole growth given a Salpeter time of $t_{\rm
  S}\simeq 180\,{\rm Myr}$ \citep[for Eddington ratios
  $L_{\rm bol}\slash L_{\rm Edd}\simeq
  0.25$;][]{Kollmeier2006}. However, \citet{White2008} also argued that
the dispersion $\sigma$ in the relationship between quasar luminosity
$L$ and dark matter halo mass $M_{\rm halo}$ must then be less than
$50$~per cent ($99$~per cent confidence) for this high $t_{\rm dc}$. A decrease in
duty cycle to $t_{\rm dc}\simeq 100$~Myr would already require an unphysically
small  amount of scatter $\sigma \lesssim 10$~per cent in the $L-M_{\rm halo}$
relation.  Thus, if our quasars emit their radiation in one continuous
episode such that $t_{\rm Q} = t_{\rm dc} \sim 1\,{\rm Myr}$, there appears
to be no easy way to reconcile our results with the clustering
measurements. Furthermore, for our short inferred lifetimes and the
assumption of a simple light curve,
it would be impossible to grow $\sim 10^9\,M_{\odot}$ black holes
in these quasar hosts without invoking super-Eddington accretion rates (Davies et al in prep.). 

One way to solve this problem would be to invoke a so-called
flickering light curve model instead of a light bulb one
\citep{Ciotti2001,Novak2011, Oppenheimer2013}.
In this picture the ultraviolet continuum emission from the quasar
fluctuates as a result of either intrinsic changes in the accretion
flow, or time variable obscuration along our line-of-sight. In
general, the response of proximity zones to flickering light curves
depends on the details of the light curve shape and the ionization
state of the ambient IGM around the quasar.  But the key point is that
it takes the IGM an equilibration time-scale $t_{\rm eq} \simeq
1\slash \Gamma_{\rm bkg}^{\rm HeII}$ to respond to changes in the
radiation field. For the sake of illustration, consider a toy model
light curve whereby quasars emit continuously as light bulbs for
$t_{\rm on} = 1\,{\rm Myr}$, but are then quenched for $t_{\rm off}
= 10\,{\rm Myr}$, and that this on/off behavior continues over
a Hubble time $t_{\rm H}$.  If \ion{He}{ii} in the IGM is highly ionized,
then because $t_{\rm off}$ is comparable to the equilibration time
$t_{\rm eq} \simeq 30\,{\rm Myr}$, gas in the IGM has
enough time to recombine to ambient IGM ionization levels during the off
periods. As a result, the lifetimes we would infer from studying the
proximity zones of active quasars would be $t_{\rm Q} \simeq t_{\rm
  on}$, which however grossly underestimates the duty cycle $t_{\rm
  dc} = (t_{\rm on}\slash t_{\rm off})t_{\rm H} = 160\,{\rm
  Myr}$ at $z\simeq 4$. This toy model suggests
that one can find a flickering light curve model which can satisfy
our proximity zone constraints and still provide a sufficiently long duty
cycle $t_{\rm dc}\sim 100\,{\rm Myr}$ required
to grow the SMBH and closer to the values deduced from clustering. Although we
note that the $z\sim 4$ quasar clustering results appear to tightly
constrain the allowed light curves, since according to \citet{White2008}
$t_{\rm dc}\sim 100\,{\rm Myr}$ would start to imply
an unphysically small  amount of scatter $\sigma$ in the relationship between quasar luminosity
and and dark matter halo mass. More careful modeling of flickering light curves
is clearly an interesting subject for future work. 
 
\section{Conclusions}
\label{sec:consl}

We have measured the \ion{He}{ii} proximity zone sizes in the spectra of six $z \simeq 4$ quasars. We performed cosmological hydrodynamical simulations, post-processed with $1$D radiative transfer algorithm to analyze these \ion{He}{ii} proximity zones. We have used a fully Bayesian MCMC formalism to compare the distribution of \ion{He}{ii} proximity zone sizes in simulations to the sizes of observed proximity zones in order to infer the quasar lifetimes, as well as the initial \ion{He}{ii} fraction in the IGM surrounding these quasars. 

Our simulations indicate that proximity zone sizes are relatively
insensitive to the \ion{He}{ii} fraction of the ambient IGM
surrounding the quasars, which is confirmed by the results of our
inference. We thus marginalize over the unknown $x_{\rm HeII}$ to
obtain constraints on quasar lifetime.  We inferred $
{\rm log}_{10} \left( t_{\rm Q} / {\rm Myr}\right)  =
-0.20^{+0.36}_{-0.45}$ for quasar SDSS~J1319$+$5202 and  $
{\rm log}_{10} \left( t_{\rm Q} / {\rm Myr}\right)  =
0.76^{+0.26}_{-0.28}$ for the HE2QS~J1630$+$0435, but
were able to put only $95$~per cent limits on the lifetime of
the remaining quasars due to large uncertainties in their systemic
redshifts. In order to mitigate the effect of redshift error, we have
also performed a joint analysis on four quasars and find $\langle
{\rm log}_{10} \left( t_{\rm Q} / {\rm Myr}\right) \rangle =
0.07^{+0.40}_{-0.55}$,
which is consistent with the two other measurements. All of our results thus seem to point to
quasar lifetimes of $t_{\rm Q} \sim 1$~Myr at $z\sim 4$. We discussed this result in the
context of other lifetime estimates at $z\gtrsim 6$ that seem to deduce a comparable value,
as well as the implication from quasar clustering that the duty cycles of $z\sim 4$ quasars
are much longer.

Unfortunately, the large uncertainties inherent in using broad emission lines in the
rest-frame UV/optical to determine quasar redshifts  significantly limit the precision with which
we can measure the lifetimes of individual quasars. An important direction for the
future would be to obtain accurate systemic redshifts of these and other \ion{He}{ii} quasars
via mm and sub-mm observations of CO and [\ion{C}{ii}] $158\mu{\rm m}$ lines arising
from cool gas reservoirs in the quasar host galaxies. Indeed, the much smaller systemic
redshift errors ($\sim 50\,{\rm km\,s^{-1}}$) would enable lifetime measurements for all
of the quasars considered here with a much smaller error of $\sim 0.10$ dex.
We also note that a large sample of $\sim 20$  \ion{He}{ii} Ly$\alpha$ forest spectra
exist at $z\sim 3$, and besides significantly
improved statistics one also benefits from much better constraints on the
\ion{He}{ii} fraction of the ambient IGM. We emphasize that the 
statistical techniques presented in this paper can
also be applied to the measurements of quasar lifetime from the
\ion{H}{i} proximity effect at $z \simeq 6$ (\citealp{Eilers2018}, Davies et al in prep).
Furthermore, it would be interesting to explore statistical methods
(for both \ion{He}{ii} and \ion{H}{i} proximity zones) which uses the entire
transmission profile
\citep{Davies2018} instead of just $R_{\rm pz}$.
Finally, it would be interesting
to perform a joint analysis of the line-of-sight
\ion{He}{ii} proximity effect and the thermal proximity effect (resulting from
\ion{He}{ii} photoelectric heating)  in the 
\ion{H}{i} Ly$\alpha$ forest \citep{Khrykin2017} at $z \simeq 4$,
which will provide additional constraining power (especially for determining $x_{\rm HeII,0}$)
as well as an independent check of our results.

\section{Acknowledgments}
\label{sec:ackn}

The authors thank Frederick Davies, Christina Eilers, Hector Hiss,
Michael Walther, and Tobias Schmidt for useful
discussions. I.S.K. acknowledges support from the grants of the
Russian Foundation for Basic Research (RFBR) No. 18-32-00798 and
No. 17-52-45063, and partial support from the grant of the Ministry
for Education and Science of the Russian Federation
(3.858.2017/4.6). G.W. acknowledges support by the Deutsches Zentrum
f\"ur Luft- und Raumfahrt (DLR) grant 50OR1720. Partly based on
observations made with the NASA/ESA Hubble Space Telescope, obtained
at the Space Telescope Science Institute, which is operated by the
Association of Universities for Research in Astronomy, Inc., under
NASA contract NAS 5-26555. Support for this work was provided by NASA
through grant number HST-AR-15014.003-A from the Space Telescope Science
Institute, which is operated by AURA, Inc., under NASA contract NAS
5-26555. These observations are associated with Programs 13013, 13875, and 14809. Archival Hubble Space Telescope data (Program 12249) were obtained from the Mikulski Archive for Space Telescopes (MAST).

\bibliographystyle{mnras}
\bibliography{References}

\bsp	
\label{lastpage}
\end{document}